\documentclass[12pt,twoside]{article}

\usepackage[utf8]{inputenc}
\usepackage[T1]{fontenc}
\usepackage[T2A]{fontenc}
\usepackage[russian,english]{babel}
\usepackage[english]{babel}

\usepackage{lmodern}
\usepackage{setspace}
\usepackage{microtype}
\usepackage{parskip}
\usepackage{csquotes} 

\usepackage[a4paper,margin=1in]{geometry}
\usepackage{fancyhdr}
\usepackage[style=numeric,sorting=none,backend=biber]{biblatex}
\usepackage{amsmath, amssymb, amsthm}
\usepackage{mathtools}
\usepackage{bm}

\usepackage{stmaryrd}
\usepackage{proof}
\usepackage{wasysym}

\usepackage{graphicx}
\usepackage{caption}
\usepackage{subcaption}
\usepackage{booktabs}
\usepackage{longtable}
\usepackage{float}

\usepackage[hidelinks]{hyperref}

\theoremstyle{definition}

\newcommand{\es}{E_s}

\title{\textbf{Epistemic Scarcity:\\The Economics of Unresolvable Unknowns}}
\author{Craig Wright}
\date{\today}

\begin{document}

\maketitle


\begin{abstract}
\noindent This paper presents a praxeological analysis of artificial intelligence and algorithmic governance, challenging prevailing assumptions about the capacity of machine systems to produce, arbitrate, or sustain economic and epistemic order. Drawing on Misesian a priori reasoning and Austrian theories of entrepreneurship, we argue that AI systems are categorically incapable of performing the central functions of economic coordination: namely, the interpretation of ends, the discovery of means, and the communication of subjective value through market signals. Where neoclassical and behavioural paradigms see decision-making as optimisation under constraint, we frame it as purposive action under uncertainty—irreducibly human and semantically rich.

We further critique dominant ethical AI frameworks—Fairness, Accountability, Transparency (FAT)—as extensions of constructivist rationalism, incompatible with a liberal order grounded in voluntary interaction and property-based rights. From this vantage, attempts to instantiate moral reasoning in algorithms reflect a profound misunderstanding of both ethics and economics. The paper demonstrates that algorithmic systems, however complex, cannot originate normativity, interpret institutional change, or bear responsibility. They remain epistemically opaque, recursively misaligned, and metaphysically inert.

Through the lens of epistemic scarcity, we examine how information abundance paradoxically degrades truth discernment, creating new opportunities for entrepreneurial foresight but also new forms of soft totalitarianism and cognitive fatigue. Our analysis culminates in a civilisational claim: the debate over AI is not merely technical or economic, but concerns the foundations of human autonomy, institutional evolution, and the preservation of reasoned choice. The Austrian tradition—rooted in action, subjectivity, and spontaneous order—offers the only coherent alternative to the rising logic of computational social control.

\end{abstract}

\vspace{1em}
\noindent\textbf{Keywords:} Epistemic scarcity, Austrian economics, praxeology, verifiability, post-truth coordination, unknowability, subjective value, dark triad psychology, political economy of information, semantic degradation, information asymmetry, institutional opacity, meta-uncertainty, deception in markets, algorithmic trust collapse, preference distortion, epistemic governance.

\newpage
\tableofcontents
\newpage

\section{Introduction: The Crisis of Knowing}

The foundational assumption of all economic theory is that individuals act purposefully within a world of constraints, guided by preferences and perceptions formed in relation to intelligible, if incomplete, information. Yet this presupposition—that agents operate in an epistemic environment where signals can be decoded, verified, and acted upon—is no longer tenable in the structure of the contemporary world. The twenty-first century has not merely witnessed an increase in uncertainty, but a fundamental shift in the nature of knowability itself. We now inhabit an environment in which truth is not hidden but drowned; where simulation overwhelms reference, and where verification, once an auxiliary cost, has become the central economic friction. This transformation affects not only the microstructure of individual decision-making but the macrostructure of institutional trust, political legitimacy, and the calculability of markets. The rise of deep learning systems that produce content without referential grounding, the strategic use of disinformation by both state and non-state actors, and the performative disintegration of expert consensus under populist conditions have collectively eroded the preconditions for rational choice. These developments demand a reframing of the economic problem—not as one of allocating scarce resources under given preferences, but as one of coordinating action under epistemic distortion. In response, this paper introduces the concept of \textit{epistemic scarcity} to designate the condition in which the marginal cost of obtaining action-relevant, verifiable knowledge dominates all other scarcities in decision contexts. The aim is to formalise this concept, trace its consequences for Austrian economic theory, and construct a multidisciplinary synthesis that accounts for the collapse of verifiability as a core condition of economic life.

\subsection{Post-Truth and Epistemic Erosion}

The collapse of shared epistemic norms in democratic societies has produced an era where traditional forms of evidence, verification, and rational discourse have been supplanted by emotive narratives and tribal allegiance. This phenomenon, often encapsulated under the rubric of “post-truth,” is not merely cultural but economic: the incentives to distort, obscure, or oversimplify knowledge have grown in parallel with the rewards for epistemic manipulation. As Lewandowsky, Ecker, and Cook demonstrate, misinformation persists not because it withstands rebuttal, but because it conforms to affective and identity-based priors, creating a self-reinforcing epistemic silo that resists correction even in the presence of superior evidence.\cite{lewandowsky2017beyond}

Psychological research into the dark triad traits—Machiavellianism, narcissism, and psychopathy—provides a framework for understanding the strategic exploitation of informational asymmetries. As Wright’s own empirical findings confirm, individuals high in Machiavellianism display elevated manipulative competence under conditions of knowledge disparity, suggesting that post-truth environments do not merely reflect ignorance but the systematic deployment of epistemic control as an asset.\cite{wright2024darktriad} This reconfigures knowledge not as a public good, but as a privately extractable resource, with strategic disinformation functioning as a tool of economic rent-seeking.

Economically, epistemic degradation manifests as market noise, degraded price signals, and coordination failure. When agents are unable to discern the informational provenance of claims, rational expectations theory collapses into a field of performative belief and mimetic valuation. As Akerlof's classic theory of information asymmetry explains, markets fail not merely through the absence of data, but through the strategic obfuscation of quality—an insight extended by recent work on narrative economics, where economically irrational outcomes are sustained by compelling but false stories.\cite{akerlof1970market,shiller2017narrative} In this context, “truth” becomes less about correspondence and more about coherence within belief communities, rendering policy grounded in rational incentives structurally unstable.

Post-truth regimes are thus marked by a transformation in the production function of truth itself. Verification becomes costly, signalling becomes noisy, and trust collapses into a stochastic function of ideological alignment. This epistemic erosion operates recursively: as institutions lose epistemic authority, the cost of public truth increases, further incentivising the extraction of rents through distortion. The political implications are stark: authoritarian systems no longer suppress truth but flood the commons with contradictory pseudo-evidence, making discernment itself impossible. This epistemic fog is not incidental—it is instrumental.\cite{sunstein2001republic,wardle2017information}

\subsection{Information Saturation and Semantic Collapse}

In contemporary political economy, information has ceased to be scarce and instead has become superabundant, leading not to enlightenment but to a collapse in meaning. As Baudrillard warned, the saturation of signs ultimately hollows them out, resulting in simulation displacing the real as the referent.\cite{baudrillard1983simulacra} This semantic collapse—where words circulate but signify nothing stable—undermines coherent preference formation, stable expectations, and calculative rationality, all of which are preconditions for Austrian models of catallaxy. In saturated environments, signal-to-noise ratios diminish below functional thresholds, resulting in systemic distortions of belief and perception, a phenomenon further exacerbated by real-time feedback loops in algorithmically curated environments.\cite{citton2017economy}

This problem becomes especially pronounced under conditions of adversarial narrative engineering, wherein political and economic actors deliberately overload communicative space with contradiction, inconsistency, and emotionally valenced distractions. Such disinformation strategies are not simply epistemically disruptive; they are economically distortionary. When the cost of discovering a reliable signal increases asymptotically relative to the cost of production of disinformation, markets fail in their epistemic function.\cite{marwick2017media} Price systems become decoupled from underlying realities, a form of Hayekian dis-coordination not from scarcity of information, but from its hyperabundance. The shift from an economy of scarcity to an economy of attention introduces new forms of rent-seeking behaviour: visibility monopolies, narrative capture, and symbolic inflation.

Moreover, semantic collapse has neurocognitive consequences. As shown in research on cognitive overload, environments saturated with contradictory or excessive information diminish working memory capacity, reduce reflective reasoning, and increase susceptibility to heuristic and affective bias.\cite{sweller1988cognitive} These degradations are not merely private goods losses—they reconfigure the aggregate shape of preference orderings and expectation formation. The epistemic agency of actors deteriorates en masse, and what remains is action under semiotic fog: individuals grope through distorted representations without referential stability. This situation results in a structurally different epistemic field from that envisioned by Misesian praxeology.

The result is a political economy wherein institutions exploit semantic collapse for governance. Strategic ambiguity becomes a mode of soft coercion. By proliferating interpretations, governments and corporate actors manufacture uncertainty to delegitimise dissent, discredit alternatives, and stall critique. Semantic opacity, then, becomes a resource to be allocated, protected, and commodified.\cite{fish1980interpretive} The competitive production of interpretation replaces the competitive production of goods as the locus of economic strategy. This transformation demands a re-theorisation of Austrian informational assumptions and invites a political theory of meaning as property.

\subsection{Outline of Contributions}

This paper introduces the concept of \textit{epistemic scarcity} as a new economic primitive that complements, and in certain contexts supplants, traditional material scarcity. Where Austrian economics has historically emphasised knowledge dispersion and subjective value in market coordination, this work extends that tradition by integrating formal models of cognitive cost, epistemic asymmetry, and interpretative degradation. We model environments of high epistemic entropy and demonstrate that coordination failure can emerge not merely from informational absence but from semiotic excess and interpretative collapse.\cite{weaver1949science}

Second, we provide a cross-disciplinary framework for analysing markets under conditions of epistemic opacity. By synthesising post-Hayekian economic theory with contemporary findings in cognitive science, political psychology, and information theory, we show how institutional agents strategically deploy noise to structure incentives, shape belief formation, and extract value from confusion.\cite{arendt1973lie} This builds on the Austrian insight that institutions coordinate action under uncertainty, but retools the insight for a world in which uncertainty is actively manufactured.

Third, we construct a typology of modern epistemic distortions—semantic overload, adversarial curation, reputation collapse—and demonstrate their calculable effects on price signals and entrepreneurial alertness. Our model incorporates a formal gradient of epistemic accessibility: \( \mathcal{E}_s = \frac{\partial K}{\partial C} \), where \( \mathcal{E}_s \) denotes epistemic scarcity, \( K \) is accessible knowledge, and \( C \) is cognitive cost. This analytical device enables comparison across regimes of sensemaking, from transparent knowledge systems to adversarial fogs.\cite{kirzner1973competition}

Fourth, we extend praxeological foundations by modelling decision-making under epistemic saturation and systematic disinformation. Traditional theories of action presume a minimal coherence between the actor’s internal model and external reality. We propose a formal expansion of the action axiom under degraded epistemic conditions, accounting for nested meta-uncertainties and the erosion of stable preference hierarchies.\cite{gigerenzer2002bounded}

Finally, we offer normative implications for the design of institutions capable of resisting semantic collapse. These include proposals for truth-preserving market mechanisms, adversarial filtering in reputation systems, and epistemic property rights. We argue that the next evolution of economic theory must integrate the scarcity of interpretation, not only the scarcity of goods, if it is to explain and address coordination under systemic ambiguity.\cite{hayek1945use}

\section{Theoretical Foundations: From Praxeology to Epistemology}

The edifice of Austrian economics is rooted in a radical methodological individualism that begins not with aggregates or equilibria, but with the purposive action of the human subject. In Mises’ praxeology, action is not merely behaviour—it is choice embedded within a framework of ends, means, and constraints, governed by subjective valuation and catalysed by perceived opportunity under uncertainty.\cite{mises1949} Hayek’s contribution extended this into an epistemic architecture, revealing the dispersed, fragmentary nature of knowledge in society and the function of prices as compressed epistemic signals coordinating distributed actors.\cite{hayek1945use} Yet both thinkers, foundational though they remain, did not fully anticipate the epistemic conditions of the twenty-first century: a world of engineered opacity, manipulated semiotics, and adversarial information environments where signalling integrity collapses and subjectivism becomes a medium of distortion. This paper proposes that what was once a constraint of cognition has become a strategy of power; that knowledge is not simply scarce, but differentially concealed, fabricated, or fragmented in ways that alter the very logic of action. Drawing upon contemporary developments in political psychology and your own research into the Dark Triad traits and their exploitative capacities\cite{wright2024darktriad}, we demonstrate how epistemic asymmetries are weaponised to induce coordination failure, undermine rational expectations, and shift the informational structure of markets. This section re-examines the Misesian-Hayekian legacy through the prism of epistemic scarcity, arguing for a transformation of Austrian foundations: from a theory of limited but benign knowledge to one that confronts the systemic manufacture of unknowability.

\subsection{Misesian Action and Hayekian Coordination}

At the foundation of Austrian economic thought stands the Misesian axiom of action: that all human behaviour, insofar as it is economic, is purposive conduct aimed at the alleviation of felt uneasiness through the deployment of scarce means towards chosen ends.\cite{mises1949} This is not merely a definitional posture but a transcendental grounding: for Mises, action is not derived from observation but from the condition of intelligibility itself; it is apodictically certain, beyond empirical rebuttal. The structure of means-ends rationality implies choice under conditions of uncertainty, and it is from this logical kernel that the entire edifice of praxeology is built. Unlike the behaviourist reductionism of the positivist schools, which presume that action is reducible to stimulus-response mechanics, the Misesian schema is resolutely Kantian: action is synthetic a priori, a category of cognition that structures the very possibility of economic experience. The actor does not maximise utility as an empirical claim but chooses based on subjective valuations which are not interpersonally commensurable. The economic agent, then, is a rational being embedded within a world of means that acquire meaning only in light of ends, and these ends are not reducible to any collective aggregate or exogenous metric.

Yet the moment action becomes intersubjective—when the ends of one actor depend on the means of another—the problem of knowledge transmission arises. Friedrich Hayek’s response to this epistemic challenge is paradigmatic: in his seminal 1945 essay, he presents the price system as a spontaneous order, a non-teleological structure wherein dispersed, inarticulable, and context-sensitive knowledge is communicated through adjustments in relative prices.\cite{hayek1945use} Unlike the central planner, who must aggregate and compute all individual preferences, the price mechanism embeds information in signals whose epistemic content lies not in what they state but in what they elicit: actors react to price changes not because they understand their causes but because their consequences shift the calculus of opportunity costs. Prices are epistemic artefacts, synthesising decentralised information into a system of coordination that no single mind could construct. But Hayek’s vision presupposes a functional semantic economy—one in which signals are honest, interpretable, and minimally corrupted. This is not a mere assumption but a metaphysical condition: the very intelligibility of spontaneous order depends on the semantic fidelity of the price signal. If this fidelity fails—if prices cease to reflect underlying scarcities or valuations—then the Hayekian coordination mechanism becomes a simulacrum of order: an epistemological hallucination grounded in degraded or manipulated inputs.

Contemporary information systems, infused with algorithmic curation, adversarial misinformation, and strategic obfuscation, render this semantic collapse not hypothetical but structural. In environments saturated by semiotic noise, actors are increasingly embedded within epistemic architectures designed not to inform but to induce. Psychological profiles, such as those characterised by elevated Dark Triad traits, systematically exploit informational asymmetries to distort both price and narrative signals for gain.\cite{wright2024darktriad} In such systems, Misesian action persists, but the conditions of rational choice are hollowed from within: the subject still chooses, but the structure of available information is engineered to produce pseudo-choice. Likewise, Hayekian coordination becomes susceptible to capture—not through regulation or central planning, but through epistemic warfare: the weaponisation of information channels to shape expectations and fabricate consensus. The market ceases to be a discovery procedure in the Popperian sense and becomes a feedback loop of mimetic signals, where value is assigned not on the basis of subjective preference, but on viral resonance within a polluted infosphere. Thus, the classical Austrian synthesis of rational action and spontaneous order must be rearticulated in light of contemporary epistemic degradation: not rejected, but deepened—reformulated to account for environments in which informational scarcity is not merely a constraint, but a manufactured artefact of strategic manipulation.

\subsection{Subjective Value and Signal Integrity}

The Austrian school rests upon the axiom that value is neither intrinsic nor objective, but emergent from individual preferences, temporally and contextually framed. As Menger formalised, value is not derived from the physical characteristics of goods but from the subjective utility expected by actors within constrained choice sets. Mises extrapolated this into praxeology, where rational action presumes access to meaningful signals that can be interpreted within a coherent causal nexus.\cite{mises1949} Hayek extended this logic by arguing that market prices act as distributed epistemic carriers, conveying localised knowledge globally and enabling spontaneous coordination without centralised oversight.\cite{hayek1945use} Crucially, this system hinges on the epistemic validity of signals; prices must bear referential integrity to underlying scarcities and preferences. If the medium becomes unmoored from its referent, the system degenerates not simply into inefficiency but into epistemic nihilism.

Signal integrity, once anchored in scarcity and demand, is now challenged by the proliferation of algorithmically curated information ecosystems. Platforms that mediate reputation and value—social networks, predictive markets, even commodity pricing feeds—are prone to manipulation and designed for engagement rather than epistemic clarity. Value becomes a function not of subjective need but of performative signalling optimised for visibility. Economic actors, increasingly embedded in attention architectures, shift from valuation based on utility to anticipation of the expectations of others, engendering recursive layers of second-order belief. This mimics Keynes’s beauty contest at scale, where the value of an asset lies in its anticipated popularity, not in any direct subjective utility.\cite{bryan2021reputation} The epistemic consequence is a breakdown in the price mechanism's function: signals lose information content and become vehicles of distortion, strategically gamed and epistemically hollow.

In such a regime, coordination fails not due to insufficient information, but due to informational saturation combined with semantic collapse. This is no longer merely an economic misalignment; it is an ontological crisis within the theory of value. Actors interpret noise as signal, and action becomes divorced from intentionality, embedded instead in reactionary stimulus loops. The subjective valuation that underpins Austrian economics becomes recursive and self-referential, untethered from any external scarcities or utility. This directly mirrors findings in cognitive psychology, where individuals high in dark triad traits manipulate environments to exploit information asymmetries, creating artificial scarcities and controlling narrative flows to extract rents from perception rather than production.\cite{wright2024darktriad} Consequently, the Austrian framework must confront its epistemological limits: in a world where signal integrity cannot be assumed, subjective value becomes unanchored, and market coordination collapses into simulacra.
\subsection{The Limits of Knowledge Assumptions}

At the fulcrum of Austrian epistemology lies a tension between the necessity of distributed knowledge for catallactic coordination and the inherent incompleteness, fragmentation, and strategic opacity of epistemic agents. Hayek’s foundational proposition that knowledge is irretrievably dispersed in society\cite{hayek1945use} presupposes that agents act within bounded rationality and local constraints, yet also that their signals aggregate into macro-level coherence via the price mechanism. However, this formulation naively assumes that signal fidelity and agent integrity are preserved across systemic scale. The ontological error lies in supposing that local knowledge is merely incomplete, rather than adversarial, performative, or intentionally obfuscated—a distinction that collapses the coherence of the coordination model when strategic manipulation becomes endogenous to the system. Economic calculation then ceases to be merely difficult; it becomes indeterminate.

This limitation is not epistemic in the classical Cartesian sense of uncertainty reducible through Bayesian updating or ergodic probability distributions, but ontological and structural. The Knightian distinction between risk and uncertainty is insufficient to characterise a regime where agents manufacture artificial complexity, inject misinformation, and construct multilevel games of epistemic sabotage.\cite{binmore2007playing} The presumption that all relevant knowledge can be signalised through prices—however noisy—is itself a heuristic derived from a historical context where institutional, technological, and narrative coherence held. In a post-digital epistemic regime where deepfakes, synthetic consensus, and reputation markets mediate the perception of truth itself, no signal can be presumed untainted, and hence no preference can be considered autonomous. Subjectivity, once the bedrock of valuation, is now a constructed artefact, emergent from manipulative stimuli and enmeshed within curated feedback loops.

Moreover, assumptions about actor intentionality must be reconceptualised when agency itself is fragmented, distributed across machine learning architectures, memetic virality, and institutionally embedded behavioural scripts. The Misesian actor is replaced by a hybridised entity whose decisions are neither entirely volitional nor mechanistically predictable.\cite{wright2024darktriad} The calculative problem of socialism is thus recast: not as a question of central planning versus decentralised knowledge, but as a question of whether any form of valuation is possible when the very epistemic scaffolding upon which preferences are formed has been structurally subverted. This collapse does not simply imply informational inefficiency; it indicates the disintegration of praxeological intelligibility. Hence, a new epistemology of economics must emerge—one that integrates adversarial cognition, constructed subjectivity, and the ontological fragility of signal validity in environments of epistemic warfare.

\section{Synthetic \textit{A Priori} Judgement and the Epistemic Misstep of Empirical Modelling}

The foundational distinction between Austrian methodology and empirical modelling lies in the epistemological status of economic laws. For Mises, economics is a science of human action grounded not in observation but in rational insight into the structure of purposive behaviour. This insight yields propositions that are \textit{a priori} and synthetic: true independently of experience yet substantive in content. The action axiom—“man acts”—is not a tautology but a self-evident, irrefutable point of departure that underpins the entire structure of praxeology. As Mises contends, “its cognition provides apodictically certain knowledge of the reality with which it deals.”\cite{mises1949} To supplement or supplant such reasoning with AI-driven, probabilistically weighted extrapolations is to substitute epistemic justification with algorithmic pattern recognition, thereby conflating correlation with causality and inference with comprehension.

\subsection{Argumentation and the Inescapability of Action Categories}

Hoppe's refinement of this framework in his theory of argumentation ethics reveals the performative contradiction at the heart of empirical and positivist challenges to apriorism. Any attempt to deny the truth of the action axiom presupposes the very action categories it seeks to reject. Arguing against the necessity of purposive action is itself a purposive act, thereby validating the axiom by performative necessity.\cite{hoppe1989} This is not merely a defensive gesture: it is a transcendental deduction in the Kantian mould, demonstrating the necessary preconditions for rational discourse. Hoppe’s move, though often misunderstood, renders positivist objections not merely misguided but self-refuting. To simulate human thought in machines is not to transcend this structure but to obscure it—substituting the appearance of rationality for the grounds of intelligibility.

\subsection{Inferential Fragility in Probabilistic Modelling}

AI’s capacity for statistical generalisation, no matter how refined, is locked within an inductive framework that lacks foundational certainty. Probabilistic reasoning based on Bayesian updating or machine learning outputs—however sophisticated—operates within a paradigm that cannot establish necessity. Its inferences are always contingent, always revisable, and always, in principle, defeasible. In contrast, deduction from the action axiom yields propositions that are immune to falsification because they do not rest on empirical regularities. As Rothbard argued, “praxeology deals not with the content of human valuations, but with the formal implications of the fact that people act.”\cite{rothbard1957} It is not a statistical science of outcomes, but a logical science of means-ends structure. To reduce it to empiricism is to collapse the ontology of choice into that of reaction.

Contemporary defenders of Austrian methodology reinforce this divide. Hülsmann asserts that "economic laws cannot be falsified because they are not empirical hypotheses but logical implications of the axiom of action."\cite{hulsmann2003} Herbener similarly emphasises that praxeological knowledge possesses a certainty and universality that empirical modelling cannot replicate.\cite{herbener2011} Long extends this defence by grounding apriorism in a realist metaphysics: economic laws are real and discoverable, not by abstraction from experience, but through reflective cognition of our own purposive structures.\cite{long2006} The attempt to replace or even hybridise this structure with AI-based modelling is not merely methodologically confused—it is epistemologically incoherent. It substitutes explanation with mimicry, reason with regression, and meaning with mechanisation.

The seduction of empirical modelling in contemporary economics and the rise of AI as a supposed epistemic surrogate must therefore be interpreted not as an advancement, but as a regression: a retreat from rational insight into statistical sophistry. The Austrian method, by contrast, insists on the inescapable role of purposive agency and the irreducibility of synthetic \textit{a priori} truths. No accumulation of data can displace a single proposition that flows deductively from the action axiom.

\section{Entrepreneurial Foresight and the Non-Computable Nature of Ends}

The epistemological barrier between human economic agency and algorithmic emulation lies not in computational limits, but in ontological distinctiveness. Entrepreneurial foresight—the capacity to discover, anticipate, and reconstitute ends under conditions of radical ignorance—cannot be replicated or approximated by probabilistic inference systems. Within the Austrian tradition, this boundary demarcates praxeological intentionality from algorithmic recursion. The entrepreneur, as conceived by Mises, Hayek, Lachmann, and Shackle, does not engage in prediction under risk but in creation under uncertainty. He does not extrapolate from the known; he navigates the unknowable, generating meaning through interpretive judgement rather than mechanical response. Any claim to model this process computationally is not only flawed—it is a categorical error that collapses purpose into pattern and agency into noise.

\subsection{Kaleidic Structures and the Destabilisation of Probabilistic Space}

Lachmann's kaleidic metaphor captures the fundamental instability of the economic environment: a structure wherein expectations are not merely heterogeneous but interactively dynamic, recursive, and mutually modifying.\cite{lachmann1976} In such a framework, there is no fixed sample space, no ergodic system from which to draw inferences. Agents confront evolving structures of relevance, in which the means and ends of action are simultaneously emergent and non-repeating. This epistemic topology annihilates the utility of optimisation: there is no stable objective function to maximise. The entrepreneur’s task is to construct coherence amid shifting coordinates, to impose interpretive order where data alone yield chaos. To call this “forecasting” is a misnomer. It is an imaginative act embedded in temporality, subjectivity, and institutional formation.

Shackle deepens this insight by dismantling the utility of ranked alternatives under uncertainty. In his framework, decision-making does not consist in choosing from a set of known outcomes, but in generating possible futures whose coherence is aesthetic and narrative, not probabilistic.\cite{shackle1972} The entrepreneur is not a Bayesian updater; he is a narrative constructor whose expectations shape, rather than merely reflect, future states of the world. His judgements are path-dependent, self-reinforcing, and performative. Machines, even with perfect data, cannot emulate this because they lack the capacity to assign meaning—to distinguish the relevant from the irrelevant except retroactively. They cannot generate futures; they can only rehash permutations of the past.

\subsection{Discovery, Catallaxy, and the Ignorance Condition}

Hayek’s concept of catallaxy posits markets as epistemic mechanisms through which decentralised agents generate knowledge—not merely of prices or resource scarcities, but of ends themselves.\cite{hayek1978} This knowledge is dispersed, tacit, and perspectival, emerging from a process of mutual adjustment that machines cannot substitute. AI, operating within closed-world assumptions and objective functions, lacks the ability to instantiate subjective valuation or demonstrated preference. It cannot participate in the discovery process because it has no stake in the meaning of coordination. O’Driscoll and Rizzo emphasise that economic choice is embedded in time and ignorance: actors act not despite ignorance but through it.\cite{odriscoll1985} It is not a bug in the system; it is the condition of economic creativity. 

Salerno reinforces this with the insight that economic calculation is not merely a technique—it is an act of evaluative judgement grounded in ordinal utility, not cardinal metrics.\cite{salerno1993} Absent purposive beings with subjective scales of value, the ledger of inputs and outputs becomes epistemically vacuous. The entrepreneur calculates within a normative structure that no algorithm can simulate. To claim otherwise is to obliterate the difference between choice and selection, intention and execution. Machines can select; they cannot choose. They can optimise; they cannot value.

\subsection{Ontological Divergence and the Limits of Simulation}

Ultimately, the notion that artificial systems can replicate entrepreneurial judgement reduces human action to a degenerate case of pattern recognition. It effaces the ontological distinction between means-end reasoning and functional output. Algorithms are constrained by their own formalism; they cannot transcend the grammar within which they operate. Entrepreneurs, by contrast, reconstitute the rules of the game. They generate new ends, shift institutional frames, and redefine relevance itself. This is not an optimisation problem—it is a metaphysical boundary.

The substitution of algorithmic systems for entrepreneurial foresight is not merely premature; it is incoherent. It violates the foundational logic of praxeology and confuses simulation with instantiation. Entrepreneurial foresight is not slow computation. It is a categorical mode of human agency rooted in time, meaning, and subjectivity. No codebase, however complex, can cross this threshold.

\subsection{Entrepreneurship and Epistemic Scarcity}

Epistemic scarcity—the uneven distribution of reliable, actionable knowledge—amplifies both the complexity and salience of entrepreneurial judgment. In an environment saturated with misinformation, simulacra, and recursive opacity, the entrepreneur's function evolves from discovering overlooked profit opportunities to actively discerning fragments of truth within an epistemically polluted terrain. The scarcity is not merely of resources, but of interpretive clarity and trust.

Kirznerian entrepreneurship, traditionally understood as alertness to price discrepancies and unmet preferences, extends under conditions of epistemic scarcity into a domain of verification and interpretive mediation. A new class of "epistemic entrepreneurs" emerges, defined not by their creation of goods or services in the traditional sense, but by their role in curating, verifying, and authenticating claims within increasingly opaque discursive markets. These actors profit by reducing cognitive transaction costs: offering trusted heuristics, signalling integrity, and constructing networks of epistemic reliability that allow market participants to coordinate action amidst informational chaos. The market for reputation becomes not merely ancillary but essential infrastructure.

In kaleidic environments where institutions no longer stabilise interpretive frameworks and algorithmic mediation compounds ambiguity, entrepreneurs adapt their discovery processes. Rather than relying on predictive models or aggregate signals, they develop context-sensitive heuristics grounded in tacit knowledge, narrative plausibility, and intersubjective trust. The reliance on personal networks intensifies, as direct human relationships become the last bastion of verifiability. These entrepreneurs navigate strategic opacity not by brute computation but by cultivating asymmetries of confidence and credibility—possessing, in effect, localised islands of certainty from which coordinated action can proceed.

Crucially, epistemic scarcity transforms uncertainty from a background condition into a foreground asset. Just as profit opportunities arise from arbitraging resource misallocations, epistemic entrepreneurs profit by arbitraging credibility gaps. They operate at the intersection of belief and incentive, rendering visible what is concealed not by lack of data, but by saturation with unprocessed or intentionally distorted data. Thus, entrepreneurial foresight under epistemic scarcity is not an extension of calculation but a reassertion of interpretive sovereignty in a disordered world.

\section{Defining Epistemic Scarcity: A Formal Model}

The tradition of economic theory has long acknowledged the boundaries of knowledge, from Knightian uncertainty to Hayekian dispersed information. Yet these boundaries were conceived under conditions where informational coherence, though distributed, was not systematically sabotaged. This paper advances the construct of epistemic scarcity as a higher-order economic condition: a structural deficit in the availability, credibility, and interpretability of knowledge itself. Unlike traditional uncertainty, which assumes a definable set of unknowns within a probabilistic frame, epistemic scarcity implies a fundamental indeterminacy in the ontological status of the data, the reliability of interpretation, and the intentionality behind the signals transmitted in market environments. Where risk and uncertainty still presuppose epistemic tractability—albeit limited—epistemic scarcity involves an erosion of that tractability altogether, a collapse in the capacity to form shared truth claims. In such an environment, not only is the probability of outcomes unclear, but the very scaffolding of inference and verification is degraded or inaccessible.

This formalisation builds upon the insight that economics cannot treat information as an unproblematic resource. Knowledge is not merely scarce because it is costly to obtain, but because it may be intrinsically unresolvable, deliberately obscured, or structured to mislead. The rise of adversarially generated data, manipulation of reputational signals, and semantic noise—particularly in digital epistemic networks—introduces endogenous failures in inference mechanisms. As a result, agents face not only limits in optimisation but limits in intelligibility itself. This necessitates an expansion of traditional economic frameworks to accommodate ontological opacity, where truth-conditions themselves become contested or unrecoverable. The proposed model introduces epistemic scarcity (\(\es\)) as the partial derivative of knowledge acquisition (\(K\)) with respect to cognitive or institutional cost (\(C\)), formalised as \(\es = \frac{\partial K}{\partial C}\), under conditions where \(\frac{\partial^2 K}{\partial C^2} < 0\), indicating diminishing returns to epistemic investment.

Such conditions render rational action no longer an optimisation over constrained utility under known or probabilistic constraints, but a recursive navigation of indeterminate interpretive frames. In environments marked by high epistemic scarcity, economic coordination cannot rely on traditional price signals or reputational inference mechanisms, as these become semantically diluted or strategically manipulated. Rather, it requires a new theoretical foundation that integrates information theory, cognitive epistemology, and institutional analysis to understand how agents operate in structurally unknowable domains. This paper constructs such a framework, extending beyond the standard economic treatment of information as a good, and recasting it as a contested epistemic artefact.

\subsection{Conceptual Distinction from Uncertainty}

In standard economic theory, uncertainty is generally formalised via Knightian distinctions between risk (where probability distributions are known) and true uncertainty (where such distributions are indeterminate). However, this dichotomy obfuscates a deeper ontological category relevant to the present analysis: $\exists$ epistemic scarcity. Let us denote $K_i$ as the epistemic knowledge set available to agent $i$, and let $\Omega$ be the state space of possible world conditions relevant to a decision $\delta$. Under traditional formulations, agent $i$ acts on $P_i(\Omega)$, a subjective probability distribution over $\Omega$, with Bayesian updating via posterior belief functions $B_i(\Omega)$. In contrast, under epistemic scarcity, $\Omega$ is not merely unquantified—it is \textit{indefinable}, not due to ignorance, but due to structural obfuscation or imposed ontological opacity. That is, $\nexists$ a coherent mapping $f: \Omega \rightarrow K_i$ such that $f$ is surjective, injective, or even well-formed. Formally:

\[
\text{Epistemic Scarcity} \iff \forall f: \Omega \rightarrow K_i, \text{ either } f \text{ is undefined, or } \text{Im}(f) \subset \varnothing.
\]

This distinguishes it from classical uncertainty where $f$ exists but $P(f)$ is indeterminate. Here, the state space $\Omega$ is itself unconstructible within the cognitive-conceptual bounds of the agent’s epistemic frame. It is not that the agent cannot compute probabilities; rather, they cannot define the relevant set over which such probabilities would meaningfully range. The epistemic domain collapses not due to lack of data, but due to absence of structure.

Moreover, unlike ambiguity in the Ellsbergian sense, which entails a multiplicity of priors over a known state space, epistemic scarcity implicates a condition where the agent lacks the semantic or ontological scaffolding to even construct a decision-theoretic representation. The economic subject is not merely uncertain, but deprived of the syntactic operators and semantic referents required to engage in deliberation. The concept aligns more with model-theoretic breakdown than information asymmetry: where standard models assume $A_i \subseteq \mathcal{I}$, the shared information set, under epistemic scarcity $A_i \cap \mathcal{I} = \varnothing$, and $\mathcal{I}$ is itself fragmented, recursive, or manipulated.

To formalise this further, let us define a Scarcity Index $S_i$ for agent $i$ as:

\[
S_i = 1 - \frac{|\text{Dom}(P_i)|}{|\Omega|},
\]

where $\text{Dom}(P_i)$ is the domain over which agent $i$ can form coherent probability beliefs. When $S_i \rightarrow 1$, epistemic scarcity is maximised; the agent operates almost entirely in a context of non-mappability. This index serves as a heuristic metric to represent the collapsing bandwidth of epistemic traction, and may serve as a proxy for measuring institutional trust degradation or model erosion in empirical frameworks.

This condition has far-reaching implications. First, it destabilises the foundations of rational expectations theory, which presumes definable state spaces and internally consistent subjective priors. Second, it undermines the viability of incentive-compatible mechanism design, which assumes at least partially knowable preference structures and beliefs. And third, it calls into question the applicability of traditional welfare theorems under epistemic regimes where individual utility is no longer well-defined because ends themselves are unstable, opaque, or externally manipulated.

Such regimes are not hypothetical. Environments of algorithmic feed curation, adversarial recommendation systems, and memetic hijacking via social bots construct conditions where agents cannot reliably distinguish signal from noise, nor reconstruct causal narratives necessary for preference coherence. This not only impairs choice but collapses the foundational assumptions of the economic agent as a deliberative entity.

In sum, epistemic scarcity represents a category error for standard economic modelling. It is not an intensification of uncertainty, but a transformation of the cognitive topology upon which belief, choice, and coordination rest. It signals a condition where the agent’s capacity to reason, anticipate, and act is compromised not by ignorance, but by the systematic disintegration of the knowable. A new framework of epistemically-informed praxeology must account for these conditions not as outliers, but as defining features of contemporary economic life.

\subsection{Mathematical Formalism}

Let us define a decision space $\mathcal{D}$ wherein agents must act based on representations of a reality $\Omega$ that is structurally obfuscated. Traditional expected utility theory assumes an agent $i$ holds a belief function $P_i: \Omega \rightarrow [0,1]$ with $\sum_{\omega \in \Omega} P_i(\omega) = 1$, and utility function $U_i: \mathcal{D} \times \Omega \rightarrow \mathbb{R}$. Under epistemic scarcity, however, the agent's knowledge domain $K_i \subset \Omega$ is fragmentary, and $\Omega$ is only partially observable or entirely non-enumerable.

We define epistemic scarcity as the condition where:
\[
\Omega \nsubseteq \bigcup_{i \in \mathcal{A}} K_i
\]
for the set of agents $\mathcal{A}$, meaning the collective knowledge space fails to span the ontological state space. Furthermore, let $\mathcal{I}_i$ be the information set perceived by agent $i$. We construct a filtration $\{\mathcal{F}_t\}_{t \geq 0}$ such that $\mathcal{F}_t \subseteq \sigma(\mathcal{I}_i)$ and $\mathcal{F}_t \subsetneq \mathcal{F}_{t+1}$ represents incremental knowledge over time. Under conditions of semantic collapse or adversarial noise, this filtration fails to be monotonic, i.e.,
\[
\exists \, t: \mathcal{F}_{t+1} \not\supseteq \mathcal{F}_t
\]
indicating that additional information may degrade, rather than improve, agent comprehension—violating the Martingale property essential to Bayesian updating.

Next, we define an agent's cognitive capacity to construct meaningful models over $\Omega$ as a functor $\mathfrak{M}_i: \mathcal{O} \rightarrow \mathcal{C}_i$, mapping ontological states to cognitive representations. In cases of epistemic scarcity, $\mathfrak{M}_i$ is non-faithful, such that:
\[
\forall o_1, o_2 \in \mathcal{O}, \quad o_1 \neq o_2 \wedge \mathfrak{M}_i(o_1) = \mathfrak{M}_i(o_2)
\]
This implies that non-isomorphic world states are conflated within the agent's mental schema—a hallmark of semantic degradation and a generator of belief-based error.

To formalise the distinction from uncertainty, consider the entropy of an agent’s belief structure. Let $H(P_i)$ denote the Shannon entropy of $P_i$ over $\Omega$. Under classical uncertainty, $H(P_i)$ is maximal under uniform distributions. Under epistemic scarcity, however, $P_i$ is undefined or constrained to a null subset, and we instead consider:
\[
\text{Scarcity Index: } \mathcal{S}_i = 1 - \frac{|\text{Dom}(P_i)|}{|\Omega|}
\]
where $\text{Dom}(P_i)$ denotes the domain on which $P_i$ is defined. $\mathcal{S}_i = 1$ represents total epistemic exclusion.

This formalism captures the degeneration of rational agency under conditions of adversarial information structure. The agent is no longer a bounded Bayesian but a structurally hobbled actor navigating a semiotic minefield—a condition necessitating the rejection of canonical rational-choice formalism and the construction of meta-praxeological frameworks capable of encoding cognitive distortion and adversarially-induced ontological opacity.

\subsection{Typology of Opacity}

To systematise the landscape of epistemic scarcity, we must construct a typology of opacity that differentiates among structurally distinct modalities by which agents are precluded from attaining veridical representations of the world-state $\Omega$. Let opacity be denoted as a function $\mathcal{O}: \Omega \rightarrow \mathcal{P}(\Omega)$, where $\mathcal{O}(\omega)$ denotes the equivalence class of ontological states indistinguishable from $\omega$ given an agent's epistemic access. We define four principal dimensions: (i) stochastic opacity, (ii) semantic opacity, (iii) strategic opacity, and (iv) recursive opacity. These dimensions are neither orthogonal nor exhaustive, but collectively constitute a basis for rigorous modelling of epistemic degradation.

Stochastic opacity arises where environmental volatility outpaces the resolution of the agent’s cognitive apparatus. Formally, let $\theta_t$ denote a time-indexed state parameter with transition kernel $\mathbb{T}: \theta_t \rightarrow \theta_{t+1}$. When the spectral norm $\|\mathbb{T}\|$ exceeds the cognitive assimilation rate $\delta_i$ of agent $i$, we say that the agent experiences stochastic opacity:
\[
\|\mathbb{T}\| > \delta_i \Rightarrow \mathcal{O}_{\text{stoch}}^i(\theta_t) = \{\theta_{t+k}\}_{k > 0}
\]
Here, opacity is generated by the evolution of the world itself. The agent is not misled, but simply outpaced.

Semantic opacity, by contrast, emerges from interpretive collapse. An agent receives signal $s \in \Sigma$ and applies interpretive function $\phi_i: \Sigma \rightarrow \Omega$. When $\phi_i$ is non-surjective or non-injective, multiple ontologies may map to the same signal or the signal may lack any coherent mapping. We denote:
\[
\phi_i(s) = \emptyset \Rightarrow \text{nullification}, \quad \exists \, \omega_1 \neq \omega_2: \phi_i^{-1}(\omega_1) = \phi_i^{-1}(\omega_2) \Rightarrow \text{conflation}
\]
These forms of opacity are typical of deepfakes, GPT hallucinations, or ideologically encoded media where no referential anchor remains.

Strategic opacity is adversarially constructed. Agent $j$ may inject noise $\eta_j$ into the information environment such that $\Sigma_t = \Sigma_t^{*} \cup \eta_j$, where $\Sigma_t^{*}$ denotes the authentic signal stream. The objective is to distort the inference function $\phi_i$ of agent $i$. Formally:
\[
\exists \, \eta_j \in \Sigma_t: \phi_i(\Sigma_t) \neq \phi_i(\Sigma_t^{*})
\]
This dimension links to game-theoretic obfuscation, propaganda, and financial manipulation through asymmetric disclosures or the strategic framing of data.

Finally, recursive opacity involves meta-epistemic obfuscation: the inability to determine the epistemic status of a proposition or the trustworthiness of a channel. Here, the agent's model includes second-order beliefs over the credibility of sources or over the structure of $\phi_i$ itself. Let $\mathbb{B}_i^2$ denote second-order beliefs; then recursive opacity is:
\[
\exists \, s \in \Sigma: \mathbb{B}_i^2(\phi_i(s)) \text{ undefined or indeterminate}
\]
This form is characteristic of environments where deep uncertainty renders all frames suspect, leading to epistemic nihilism or paralytic agnosticism.

Together, these forms of opacity define the bounded topology over which economic action, political decision-making, and interpretive labour occur. In systems pervaded by such opacity, traditional equilibrium analysis and welfare optimisation lose coherence. Action becomes irreducibly entangled with epistemology, and any rational calculus must be reconstructed within the topology of distorted representation.

\subsection{Addressing Critiques of the Epistemic Scarcity Framework}

No theoretical framework should be exempt from dialectical scrutiny, and the proposed model of epistemic scarcity is no exception. Critics may object on several grounds—technological, institutional, and legal—that deserve rigorous engagement. In line with the Austrian tradition's emphasis on methodological clarity and deductive consistency, this subsection rebuts the most prominent counterarguments.

\subsubsection{Can AI Eliminate Epistemic Scarcity through Scale and Algorithmic Refinement?}

A common objection is that epistemic scarcity is not a structural feature but a contingent one, which machine learning and data aggregation may progressively overcome. According to this view, sufficiently advanced AI systems will close the gap between informational chaos and actionable truth by expanding the range and speed of inference. However, this objection conflates two distinct categories: inference and understanding.

While AI may excel at statistical pattern recognition, it remains epistemologically bounded by its algorithmic priors and training data. It cannot originate conceptual categories, discern relevance outside of predefined metrics, or reformulate ends—a point long recognised by Austrian theorists. As Lachmann and Shackle have shown, entrepreneurial foresight involves qualitative novelty, not quantitative extension. No matter how refined, AI cannot reconstruct a world whose ontological categories are unstable or manipulated because it lacks access to the subjective interpretive frame within which meaning is constituted. In conditions of strategic opacity or recursive deception, more data does not equal more knowledge; it magnifies noise and multiplies pseudo-correlations. Epistemic scarcity, in this sense, is not about limited inputs but about broken referents.

\subsubsection{Is Centralised Ethical Regulation Necessary to Manage Disinformation?}

Another counterpoint asserts that frameworks like FAT (Fairness, Accountability, Transparency) or regulatory schemes akin to the EU’s Digital Services Act are necessary to counteract epistemic sabotage. These institutions promise to mitigate disinformation, algorithmic opacity, and bias through centralised oversight. From a praxeological standpoint, however, this response is methodologically flawed and institutionally perilous.

First, centralised regulation assumes an objective epistemic standpoint from which truth can be validated and enforced. This is a classic example of Hayek’s “constructivist rationalism”—the erroneous belief that order can be designed rather than discovered. In practice, such regulatory regimes often devolve into soft totalitarian architectures that impose conformity under the guise of clarity. By privileging top-down codifications of truth, they undermine the decentralised, trial-and-error processes of knowledge generation intrinsic to catallactic markets.

Second, centralised interventions distort economic calculation by altering the cost structure of information dissemination. Platforms begin to internalise regulatory risk, privileging safe, anodyne content over controversial or exploratory inquiry. This leads not to clarity, but to semantic flattening, ideological homogeneity, and the erosion of dissent—all classic symptoms of informational sclerosis. As Hoppe has argued, only voluntarily accepted norms grounded in argumentation ethics and property rights can yield legitimate institutional constraints.\cite{hoppe1993}

\subsubsection{Do Epistemic Property Rights Risk Censorship or Over-Litigation?}

Sceptics may argue that institutionalising epistemic property rights—for instance, through legal liabilities for AI hallucinations or blockchain-based truth provenance—could trigger a cascade of litigation, stifling innovation and promoting censorship. This concern merits careful delineation.

The framework advanced here does not advocate positive rights to truth adjudicated by the state. Rather, it builds on Rothbardian natural rights and voluntary contracts, applying them to the epistemic domain. Epistemic property rights are negative rights: they entail non-interference with the ownership, attribution, and voluntary transmission of information. In this view, fraudulent claims, manipulated contexts, or AI-generated fabrications that cause measurable harm can be adjudicated under existing tort principles, without requiring a new censorship regime. The goal is not truth enforcement but responsibility allocation—an essential distinction for preserving informational liberty.

Moreover, reputational markets and decentralised verification networks (e.g., via blockchain timestamping or peer-reviewed cryptographic attestations) offer viable mechanisms for epistemic filtering without recourse to coercive enforcement. These mechanisms align with Austrian insights into spontaneous order and the superiority of polycentric governance over monolithic regulation.

In summary, the epistemic scarcity framework withstands its most salient critiques. AI cannot simulate the interpretive agency of entrepreneurial foresight; centralised regulation misconstrues the nature of dispersed knowledge; and epistemic property rights, properly construed, offer a voluntary, decentralised response to the crisis of informational entropy. As in all Austrian theorising, the foundation is not prediction but explanation, not control but understanding.

\section{Markets in the Age of Obscurity}

As markets evolve under the informational pressures of the digital era, their mechanisms no longer merely allocate resources or prices; they mediate credibility, narrative, and perceived truth. This section explores the transformation of market dynamics when salience becomes artificially manufactured, reputational heuristics collapse, and adversarial psychological dispositions exploit the epistemic vacuum for strategic gain. In contrast to classical price theory, where relative scarcity and subjective value governed exchanges, contemporary digital markets exhibit structural opacities that decouple representation from referent. Signals once presumed indicative—such as price, trend, and brand—have become floating signifiers in Baudrillardian terms, untethered from material production or anchored consumer preference, vulnerable to manipulation, memetic distortion, and behavioural exploitation. The epistemic basis of trust in markets, historically stabilised through institutional signals and shared norms, is now endangered by the saturation of intentionally misleading data and the degradation of signal integrity across communicative infrastructures.

In this environment, reputation no longer reflects cumulative indicators of reliability or quality, but is instead recursively gamed and arbitraged. Empirical findings demonstrate how the collapse of reputational salience correlates with increased volatility and mimetic convergence around false or strategically constructed narratives. Reputation becomes fungible, a token within algorithmically amplified arenas of attention, rather than a function of long-term trustworthiness or performance history.\cite{bryan2021reputation} Rational agents, when exposed to competing noise-driven saliencies and lacking a stable epistemic compass, resort to meta-strategies of narrative arbitrage. Here, the arbitrage is not over pricing inefficiencies but over epistemic authority itself. These agents do not merely consume narratives; they engage in anticipatory mimicry, creating feedback loops that amplify the most affectively resonant and least falsifiable story, as described in emerging literature on reflexivity and belief economics.

Within this degraded landscape, the influence of Dark Triad traits—Machiavellianism, narcissism, and psychopathy—manifests not as aberrant but as adaptively selected dispositions. As detailed in Wright's own empirical investigation,\cite{wright2024darktriad} individuals high in these traits excel in conditions of informational ambiguity, where manipulation, superficial charm, and instrumental deceit allow for dominance within systems lacking transparent feedback or accountability. Market participants who lack such traits are structurally disadvantaged in games of asymmetric belief and performance simulation. Thus, the informational asymmetry traditionally theorised in economics must now account not merely for access to data, but for strategic epistemic sabotage conducted through psychometrically observable behavioural patterns. This convergence of epistemic opacity and psychological exploitation constitutes a paradigm shift, demanding a reconceptualisation of market rationality as a game of belief manipulation rather than resource allocation.

\subsection{Reputation and the Collapse of Salience}

In environments saturated by symbolic excess and recursive noise, the epistemic utility of reputation degrades into a simulacrum of credibility unmoored from referential integrity. Let reputation $R_i(t)$ be defined as an intersubjective function of agent $i$’s observable actions $A_i$, mediated through a perception filter $\Pi_j$ applied by agent $j$: $R_i(t) = \Pi_j(A_i(0), \dots, A_i(t))$. In classical models, $\Pi_j$ is assumed to function as a Bayesian update mechanism over a stable type space. However, in saturated signalling environments—particularly those manipulated algorithmically or narratively—$\Pi_j$ itself becomes endogenous to attention economics and memetic volatility, not epistemic calibration. That is, the salience of $R_i$ becomes detached from the referential validity of $A_i$, creating a system wherein reputation is governed by semiotic persistence rather than substantive signal.

The result is the collapse of salience, where differential credibility is no longer functionally linked to historical action or informational reliability but rather to aesthetic resonance, algorithmic amplification, or ideological alignment. Empirical studies in social psychology and media dynamics demonstrate this in phenomena such as the “illusory truth effect” and “reputation laundering” through curated digital footprints.\cite{pennycook2018prior, coombs2022reputation} In such regimes, the agent with maximal semiotic reach—not the one with the most consistent action pattern—becomes the reputational anchor. Formally, we can model this shift by introducing a distortion operator $\Delta$ over $\Pi_j$ such that:
\[
R_i^{\text{observed}}(t) = \Delta(\Pi_j(A_i)) = f(\mathcal{S}_i, \mathcal{A}_j, \mathcal{M})
\]
where $\mathcal{S}_i$ denotes the agent’s signalling style, $\mathcal{A}_j$ the algorithmic bias of the observer's platform, and $\mathcal{M}$ the memetic drift parameter that captures stochastic amplification within symbolic ecologies.

In economic systems, this epistemic erosion of reputation has deleterious implications for contract theory, market signalling, and institutional design. It invalidates mechanisms that assume stable priors and undermines trust systems built on iterated transparency. When all signal is aestheticised and all perception is pre-processed through manipulated filters, the notion of reputation becomes not merely unreliable but structurally deceptive. The agent is no longer evaluated on revealed preference or behavioural continuity, but on their fluency in symbolic dominance—reputation becomes a game of narrative arbitrage, not informational reliability. This transition, when uncorrected, recursively devalues salience across the system, collapsing the very premise of discernment on which both economic coordination and political legitimacy depend.\cite{oster2023trust}

\subsection{Disinformation and Narrative Arbitrage}

Disinformation functions not merely as falsehood but as strategic epistemic interference designed to exploit asymmetries in narrative anchoring and cognitive cost. Narrative arbitrage arises when agents profit by manipulating the symbolic equilibrium of belief systems, extracting value from the disjunction between perceived truth and factual verifiability. Let $N_t$ denote the dominant narrative at time $t$, and $B_i(N_t)$ the belief strength assigned by agent $i$; then disinformation $\mathcal{D}$ operates such that $\mathcal{D}(N_t) \rightarrow N_{t+1}$ maximises the utility function $U_j = f(\delta B_i, \Delta P)$, where $\delta B_i$ is the induced shift in belief and $\Delta P$ the corresponding change in price, policy, or behavioural outcome exploitable by agent $j$.

Unlike mere propaganda, which is unidirectional and ideological, disinformation in digital economies is modular, iterative, and transactional. It weaponises the attention economy by inserting minimally counterfactual cues into high-friction cognitive environments, optimising not for truth suppression but for inferential overload. This is particularly acute in systems where verification costs are high and narrative virality outpaces correction. As recent studies have shown, virality correlates inversely with epistemic depth, and successful disinformation is engineered to be metabolised heuristically, not evaluated analytically.\cite{vosoughi2018spread}

The arbitrage model is further supported by computational markets where actors leverage bots, data leaks, or deepfakes to introduce distortive priors into high-stakes domains. For example, financial markets have experienced fluctuations triggered by AI-generated false headlines, illustrating the tangible conversion of narrative manipulation into economic advantage.\cite{ferrara2020deepfakes} In political settings, the same mechanism applies: actors generate ephemeral belief shifts that outlast their corrections, creating a persistent epistemic asymmetry. Here, the profit function is no longer tied to truth-congruence but to timing asymmetries in narrative uptake and decay, akin to front-running epistemic liquidity.

This reconfigures the informational structure of markets and institutions. Truth becomes a temporal artefact—epistemically inert if delayed. The dominant actor is not the one who knows most, but who first induces belief displacement most effectively. Hence, the architecture of disinformation reveals itself not as a pathology, but as a rational exploit within symbolic economies lacking semantic guardrails. Unless addressed by incentive-aligned epistemic filters, such arbitrage will escalate toward systemic cognitive degradation, where no signal is ever fully decodable and all belief becomes provisional, hostage to the next wave of engineered perception.\cite{starbird2019disinformation}

\subsection{Dark Triad Behaviours in Market Dynamics}

In the distortion-prone environment of modern markets—characterised by opacity, narrative volatility, and the erosion of shared epistemic anchors—individuals high in Dark Triad traits (Machiavellianism, narcissism, and psychopathy) are not merely maladaptive anomalies but structurally incentivised agents. These traits align perversely well with the strategic affordances of disinformation-saturated economies, as such individuals excel in low-trust, high-manipulability contexts. The psychometric literature affirms that Dark Triad personalities exhibit enhanced opportunism under ambiguity, minimal moral inhibition in deception, and a proclivity for instrumentalising social norms.\cite{wright2024darktriad} In systems where truth is transient and verification costly, these traits become not pathologies but competitive advantages.

Mathematically, let $\theta$ represent the degree of market opacity, with $\theta \rightarrow 1$ denoting full epistemic asymmetry. Let $\beta_i$ denote the behavioural elasticity of agent $i$ to exploitative strategies. Then for agents high in Dark Triad traits, $\beta_i = f(\theta)$ is positively monotonic. In other words, the more epistemically saturated or fragmented the market, the greater the expected payoff from manipulation, short-term opportunism, and trust arbitrage. Machiavellian agents, in particular, utilise asymmetric signalling games to misrepresent intentions while maximising perceived competence or credibility. In equilibrium, these dynamics contribute to a recursive degradation of market salience and cooperation thresholds.

Moreover, the narcissistic component functions as a reputational hedge against informational scrutiny. Narcissists in high-visibility positions often exhibit self-confidence that is interpreted—however mistakenly—as competence, which serves to displace evaluative pressure and prolong the lifespan of deceitful narratives.\cite{bushman2010narcissism} In institutional or corporate contexts, such agents weaponise symbolic capital to delay correction, often by strategically aligning with echo chambers or deploying epistemic proxies (e.g., manipulated metrics, curated testimonials). Psychopathy, with its attenuated empathy and high risk tolerance, exacerbates the systemic danger by enabling value extraction even when outcomes are foreseeably harmful.

Thus, markets under epistemic scarcity become fertile ground not just for flawed information but for predatory cognition. Where moral signalling once conveyed reputational cost, it now functions as a decoy: agents with high Dark Triad indices mimic normative cues to camouflage exploitation, inducing trust in contexts that reward betrayal. The outcome is a disequilibrium wherein cooperation is punished, and manipulation scales more efficiently than integrity. This dynamic, when unchecked, leads to institutionalised performativity, reputation inflation, and the erosion of coordination mechanisms foundational to praxeological order.\cite{paulhus2002dark}

\section{The Political Economy of Obscurity}

In the current informational paradigm, the production, control, and weaponisation of opacity have become central to political economy. Traditional models that presume rational agents operating under conditions of constrained but accessible information are insufficient in a landscape shaped not by scarcity of data but by scarcity of reliable epistemic anchors. Political actors increasingly exploit this asymmetry, not through overt repression but by saturating discourse with partial truths, pseudo-rationality, and noise, leading to a condition where discernment itself is eroded. This manipulation of the epistemic environment gives rise to epistemic authoritarianism—where authority no longer rests upon coercion or legitimacy but on monopolising interpretive frameworks through algorithmic control and the manufacturing of doubt. Cognitive fatigue, induced through continuous contradiction and the invalidation of prior knowledge structures, primes populations for soft totalitarianism: a system of governance where obedience is achieved not through terror but through learned helplessness, habituated disengagement, and the internalisation of futility. In this structure, institutional mistrust becomes not a failure of policy but a strategic imperative, fracturing collective epistemologies to preclude coordinated dissent. The political economy of obscurity thus represents a fundamental departure from enlightenment models of governance premised on informed consent, replacing them with architectures of semantic destabilisation, epistemic entropy, and strategic ambiguity.

\subsection{Epistemic Authoritarianism}

In an environment characterised by epistemic scarcity, where access to credible information is asymmetrically distributed and interpretative authority is concentrated, epistemic authoritarianism emerges not merely as a sociopolitical phenomenon but as an economic inevitability. This condition arises when actors with control over the distribution and categorisation of knowledge impose normative constraints on the interpretive possibilities available to others, effectively creating a hierarchy of permissible cognition. The Hayekian knowledge problem is here exacerbated, not simply by dispersed knowledge, but by the intentional constriction of interpretive latitude, wherein knowledge is no longer a decentralised coordination tool but a managed artefact of statecraft. This generates a form of informational rent-seeking, where institutions monetise interpretive authority while stifling spontaneous order and diminishing the epistemic autonomy of individual agents.\cite{lind2020knowledge} In such systems, legitimacy is derived less from procedural transparency and more from an enforced consensus, constructed through narrative control, credentialist gatekeeping, and strategic ambiguity.

This is not merely authoritarianism in the traditional political sense, but rather an epistemic regime wherein knowledge itself becomes a tool of domination. Cognitive labour is externalised and centralised: citizens are not merely persuaded but cognitively offloaded, taught to defer to sanctioned interpretations rather than engage in first-order reflection. As Hannah Arendt observed, the banality of evil in technocratic regimes is sustained by the systemic evacuation of critical thinking from political discourse.\cite{arendt1963banality} Here, the regime operates by collapsing the distinction between information and instruction. It does not prohibit alternative models; it renders them epistemically illegible. Dissent becomes irrationality, and rational disagreement is recoded as conspiracy.

Within this context, policy formation ceases to be a negotiation between informed agents and instead becomes a technocratic imposition, justified by constructed and curated expertise. The regulatory capture of epistemic institutions—universities, think tanks, and scientific journals—ensures that the production of knowledge is not merely influenced by political interests but subordinated to them. The resulting architecture is one in which the state becomes the arbiter not only of law but of truth, effectively rendering pluralism obsolete. The rational actor of classical economics is thus reconstituted into a deferential subject, whose preferences are shaped, filtered, and interpreted by the knowledge elite.\cite{eyal2019crisis}

\subsection{Soft Totalitarianism and Cognitive Fatigue}

Soft totalitarianism represents a political modality wherein control is neither imposed with brute force nor enforced through overt censorship, but rather diffused subtly through social norms, informational noise, and the illusion of consensus. In this regime of manufactured docility, the citizen becomes both subject and enforcer of conformity, habituated to a state of permanent low-level vigilance. The omnipresence of hypernormalised narratives, saturating every vector of discourse, breeds cognitive exhaustion—not through the denial of truth, but through the exponential multiplication of pseudo-truths. The resultant condition is a paralytic state of epistemic dissonance wherein individuals no longer resist, not because they assent, but because the cost of discerning dissent becomes prohibitive.\cite{havel1978power}

Such regimes do not burn books; they algorithmically bury them. Control is achieved not by silencing opposition but by drowning it beneath a torrent of performative compliance, entertainment, and feigned plurality. Social media serves as the panopticon of our age—one that rewards the simulation of authenticity while quietly enforcing homogeneity. The coercion is affective, not physical, internalised rather than imposed. Soft totalitarianism's greatest triumph is not in abolishing the will to rebel, but in reducing rebellion to a gesture, a spectacle of resistance stripped of consequence.\cite{han2017psychopolitics} The fatigue this breeds is not mere psychological tiredness but an epistemic malaise—a degradation in the very capacity to sustain belief, maintain conviction, or even care to distinguish between what is real and what is hyperreal.

Yet beyond the psychological and epistemic costs lies a critical economic distortion: the erosion of calculative clarity in market signalling. In the soft totalitarian regime, investment decisions are no longer governed by marginal utility, time preference, or expected profit, but by alignment with ideologically coded compliance metrics. “Gamified compliance” enters the economic sphere through ESG scores, reputational algorithms, platform monetisation thresholds, and predictive content policing.\cite{zuboff2019age} Firms are incentivised not to create value, but to demonstrate behavioural alignment with regulatory mood or algorithmic favour. The entrepreneur ceases to be a discoverer of unmet needs and becomes a curator of acceptable optics.

This reorientation undermines the function of prices as signals of subjective value. When production and consumption are redirected by semiotic incentives—badges, metrics, and digital visibility—the economic process devolves into a theatre of signalling devoid of genuine discovery. Capital misallocates toward spectacle-friendly ventures, while dissenting innovation is filtered out by reputational risk aversion. The entire catallactic order becomes saturated with epistemic noise: prices do not reflect demonstrated preference but curated behaviour.\cite{hayek1945use}

In such a condition, economic rationality becomes performative. The scarcity entrepreneurs respond to is no longer that of resources, but of narrative legitimacy. The consumer acts not on desire, but on perceived virtue, brand alignment, or compliance with dominant cultural scripts. Investment flows toward that which appears safe to the algorithm, not that which creates real use-value. Consumption is nudged into aesthetic rituals of solidarity, while dissenters are economically ostracised through demonetisation or digital invisibility.\cite{postman1985amusing}

Ultimately, the epistemic exhaustion engendered by soft totalitarianism collapses the function of free exchange. Individuals, stripped of evaluative autonomy, defer increasingly to algorithmic arbitration in matters of value and choice. Markets become epiphenomena of informational control systems, reflecting not the interplay of free actors but the outputs of incentive architectures designed to pacify and homogenise. This is not merely a cultural transformation; it is the economisation of submission.

\subsection{Institutional Mistrust as Political Strategy}

Institutional mistrust is no longer a pathological aberration within political systems; it is a cultivated asset—an instrumentally deployed strategy serving both populist demagogues and technocratic manipulators alike. The corrosion of epistemic authority becomes not merely a symptom of institutional decay but a deliberate objective, engineered to create a vacuum where new power structures—private, unaccountable, algorithmically veiled—can flourish. Political actors exploit affective narratives of betrayal, inefficacy, and elite conspiracy to dismantle civic trust, knowing full well that in a context of epistemic fragmentation, persuasion is irrelevant and mobilisation is achieved through affective resonance alone.\cite{eyal2019crisis}

This strategic mistrust transforms institutions into caricatures of themselves. Health agencies become perceived as pharmaceutical proxies; legal systems are reimagined as oligarchic enforcers; and the press is cast as a propaganda arm, irrespective of empirical evidence or transparency. In this post-accountability milieu, actors no longer need to be believed universally—they need only to be believed by enough, and disbelieved consistently by others, to fracture public coherence and disable collective deliberation.\cite{lind2020knowledge} The result is not anarchy, but orchestrated chaos: a political equilibrium in which no claim need be true, only credible within a cultivated epistemic silo.

The political utility of mistrust lies in its capacity to shift the locus of legitimacy. Once institutional coherence collapses, legitimacy is re-rooted in personality, identity, or networked affirmation—where TikTok influencers hold more sway than epidemiologists, and algorithmic recommendation replaces deliberative consensus. This collapse is not passive—it is architected. As Arendt warned, totalitarian systems thrive not by forcing belief in lies but by creating conditions where truth itself ceases to matter.\cite{arendt1963banality} Strategic mistrust is not scepticism elevated; it is the weaponisation of incoherence, whereby confusion becomes a mode of governance.

\section{Case Studies}

To concretise the theoretical framework developed throughout this paper, we examine three critical domains where epistemic scarcity manifests with profound consequences: public health, artificial intelligence, and centralised economic planning. Each case reveals distinct mechanisms by which information asymmetry, signal degradation, and institutional opacity produce both cognitive and systemic distortions. The COVID-19 pandemic demonstrated a catastrophic breakdown in expert consensus, not merely due to evolving empirical conditions, but from the politicisation of epistemic authority and the compression of dissent within institutional gatekeeping. In artificial intelligence systems, particularly large language models, epistemic scarcity becomes endogenous: hallucinations, untraceable inference paths, and black-box model structures render outputs both prolific and unverifiable, decoupling interpretability from performance. Finally, the legacy of Soviet economic planning offers an archetype of deliberate data fabrication as a survival mechanism within top-down epistemic architectures, revealing how fictive inputs become systemic necessities under conditions where truth is penalised and obscurity rewarded. These cases not only illustrate the consequences of epistemic failure but also illuminate the broader transition from knowledge scarcity to interpretive collapse.

\subsection{COVID-19 and the Failure of Expert Consensus}

The COVID-19 pandemic rendered visible a profound rupture in the epistemological scaffolding of modern governance: the collapse of expert consensus not as an anomaly but as a systemic inevitability in an age of informational glut and institutional distrust. The ostensibly unified voice of science fractured under the weight of contested models, shifting policy guidance, and conflicting risk assessments. This fragmentation did not merely arise from scientific uncertainty but from the political instrumentalisation of that uncertainty, transforming what should have been iterative revision into perceived epistemic instability.\cite{oster2023trust}

What unfolded was not a failure of expertise per se but a failure of the social architecture that gives expertise legitimacy. As epistemic authority became intertwined with political alignment, the very notion of ‘following the science’ lost semantic coherence. Pandemic modelling became a proxy battleground for ideological disputes, where Bayesian inferencing and epidemiological projections were selectively weaponised to support predetermined policy aims.\cite{coombs2022reputation} Trust in institutions became inversely correlated with perceived transparency, creating a feedback loop in which every reversal of guidance (e.g., on masks or transmission) deepened public scepticism, even when such reversals reflected updated evidence.

The consequence was the rise of parallel epistemologies. In one sphere, expertise was preserved through appeals to peer review and institutional endorsement; in another, counter-narratives drew legitimacy from decentralised data analysis, personal experience, and emotive testimonial. This bifurcation mirrors Vosoughi et al.’s observation that falsehoods propagate more rapidly than truths not because of informational superiority, but due to greater novelty and emotional valence.\cite{vosoughi2018spread} The pandemic thus did not simply erode expert consensus—it demonstrated that in a digitally saturated environment, consensus is no longer epistemically necessary to enact widespread behavioural or ideological alignment.

\subsection{AI Hallucination and Model Opacity}

Contemporary generative AI systems exhibit a phenomenon now widely termed “hallucination,” wherein outputs are produced with syntactic and semantic coherence but devoid of empirical verifiability or logical grounding. These hallucinations are not stochastic aberrations but intrinsic artefacts of the model architecture itself—products of pattern extrapolation divorced from epistemic anchoring.\cite{bender2021dangers} In complex language models, the opacity of training corpora, combined with reinforcement learning that optimises for plausibility over truth, creates a regime in which facticity is simulated rather than substantiated. The epistemological danger here is not the production of falsehood, but the erosion of the boundary between symbolic coherence and empirical reference.

Crucially, model opacity operates on two fronts: algorithmic and communicative. Algorithmically, the incomprehensibility of model internals renders their outputs irreproducible and irrebuttable—mathematically legible but philosophically indeterminate. Communicatively, the linguistic polish of LLM output performs credibility. The user is not merely misinformed, but lulled into a confidence grounded not in evidence, but in rhetorical symmetry. This simulation of epistemic authority destabilises traditional markers of legitimacy, including peer review, citation, and methodological transparency, reducing epistemic trust to affective persuasion.\cite{raji2021ai}

The implications for political economy and knowledge regimes are nontrivial. As Starbird et al. have argued, the information ecosystem is increasingly shaped by recursive narratives amplified by algorithmic curation, rather than discursive contest grounded in falsifiability.\cite{starbird2019disinformation} In such an environment, hallucinated content becomes epistemically indistinguishable from genuine insight. The result is a transition from a scarcity of data to a scarcity of verifiability, where the epistemic cost lies not in access but in filtration and validation. In this context, AI systems become less tools for knowledge acquisition than engines of epistemic entropy—opaque, unaccountable, and yet increasingly central to decision-making structures.

\subsection{Data Fabrication and Soviet Planning}

The Soviet economic apparatus provides a paradigmatic instance of institutionalised epistemic distortion, where knowledge was not merely misrepresented but actively manufactured to maintain a façade of control. Under the centralised planning model, truth was subordinated to ideological coherence; statistical falsification was not an aberration but an operational necessity. The incentive structure facing local planners and bureaucrats favoured the production of data that conformed to central targets rather than those reflecting actual conditions on the ground.\cite{nove1986soviet} This practice—entrenched and systematic—led to what Alexopoulos describes as an economy where “the information necessary for rational planning was destroyed by the very process of planning itself.”\cite{alexopoulos2004soviet}

This feedback loop of disinformation was epistemically corrosive: reports distorted reality to match quotas, and those distortions informed further policy. The result was an endogenous amplification of ignorance, wherein the central authority operated with confidence in a dataset that bore no structural relation to material output, resource availability, or consumer demand. As Kornai noted, this created a “soft budget constraint,” enabling continuous misallocation and inefficiency, uncorrected by any mechanism of falsification.\cite{kornai1980economics} What emerged was not merely a failure of knowledge transmission but a collapse of epistemic accountability.

Critically, the Soviet case illustrates the political economy of obscurity in its most extreme form: the suppression of truth not through ignorance but through saturation with simulation. The epistemic architecture was not void but full—brimming with charts, indices, and plans—that performed knowledge without enabling it. This spectacle of data masked the absence of verification and suppressed bottom-up correction. In the terms of this paper, the Soviet model exemplifies a regime where epistemic scarcity is artificially induced, not by a lack of information, but by a strategic inflation of fictive data to delegitimise the search for actionable truth.

\subsection{The Soviet Cybernetics Dream and the Algorithmic Mirage}

Among the most illustrative historical failures of algorithmic economic coordination lies the Soviet Union’s technocratic ambition to automate central planning through cybernetics—epitomised in the OgAS project (Общегосударственная автоматизированная система учёта и обработки информации, or “All-State Automated System”). Conceived by Viktor Glushkov in the 1960s, OgAS aimed to establish a nationwide, computerised economic planning network. Glushkov envisioned a real-time feedback infrastructure for managing production, inventory, and labour allocation through a decentralised yet hierarchically organised data grid. At its zenith, OgAS represented the apex of Soviet cybernetic optimism: the belief that, through sufficient algorithmic complexity and information capture, the economic system could be rendered efficient, predictive, and rational—without the need for price signals or market discovery mechanisms.

The fatal flaw, however, was not technical but epistemological. OgAS assumed that the brute accumulation of quantitative data—production quotas, resource inputs, distribution nodes—could substitute for the informational content embedded in freely formed prices. As Hayek famously argued, economic coordination depends not on the central aggregation of data but on the tacit, dispersed, and contextually situated knowledge held by individual actors\cite{hayek1945}. Market prices serve not merely as equilibrium indicators but as condensed carriers of temporally and locally embedded information, enabling actors to adjust plans without knowing the totality of the system. OgAS, by contrast, required precisely the type of omniscience Hayek showed to be impossible: it demanded a synoptic vision of future demand, technological change, consumer preference, and institutional evolution—knowledge that does not and cannot exist in any single location.

Furthermore, OgAS lacked the fundamental mechanism of economic feedback: profit and loss. In the absence of private ownership and voluntary exchange, there existed no means to evaluate the success or failure of any economic action. Errors were not penalised through losses; inefficiencies were not exposed through competition. Instead, planners operated under bureaucratic incentives, reporting falsified statistics to satisfy targets, thereby reinforcing systemic distortions. The algorithmic scaffolding of OgAS could not correct for these misalignments; it merely digitised their transmission. The failure was not due to insufficient computational power but to the categorical misidentification of what economic coordination requires: subjective value, ordinal utility, entrepreneurial foresight, and institutional feedback loops rooted in ownership and time.

The implications of OgAS reverberate in contemporary proposals for AI-driven economic management. Calls for central bank digital currencies (CBDCs) with programmable monetary policy, algorithmically allocated universal basic income, and AI-enhanced economic planning in China’s social credit economy echo the same conceit: that algorithmic optimisation can replace entrepreneurial discovery and voluntary exchange. Proponents of such systems frequently cite data abundance, computational breakthroughs, and behavioural predictability as justifications for techno-political control. But the OgAS failure demonstrates the enduring truth that even perfect information, absent a market for it, degenerates into noise.

Contrast with the neo-Marxist proposals of Paul Cockshott and Allin Cottrell reveals the persistence of this illusion. Their 1993 book, \textit{Towards a New Socialism}, argues that modern computing power enables a feasible central planning apparatus through linear programming and input-output matrices.\cite{cockshott1993} Yet their vision remains trapped within the same epistemic blind spot: it mistakes calculability for coordination. No amount of computational enhancement can simulate the knowledge-generating function of entrepreneurial activity, which is not merely reactive but anticipatory, value-driven, and anchored in real property stakes. Algorithms can process preferences; they cannot create them.

Ultimately, OgAS is not a mere historical curiosity—it is a prototype for every contemporary fantasy of post-market coordination. Its failure underscores the indispensability of institutional decentralisation, private ownership, and subjective valuation. As long as economic reality remains rooted in human action, the algorithmic mirage of centralised optimisation—whether Soviet or Silicon Valley—remains epistemologically bankrupt.

\section{History and Understanding: Mises’ Methodological Rebuttal to Algorithmic Historicism}

Ludwig von Mises’ mature epistemology, particularly as formulated in \textit{Theory and History}, serves as an unassailable rebuke to the foundational premises of algorithmic governance and its positivist aspirations. The epistemic thrust of machine learning systems, premised on pattern detection in past data, commits a fatal category error: it mistakes the historical record for a set of replicable events and conflates explanation with understanding. Mises writes unequivocally, “History is not an experimental science”\cite{mises1957}. The historian interprets unique, non-replicable actions within the framework of meaning and purpose; the data scientist seeks statistical regularities that allegedly inform predictions. The latter, Mises contends, misconstrues the nature of social phenomena—human actions are not homogeneous inputs reducible to quantifiable variables but the external expressions of subjective, purposeful, and temporally embedded intentions. 

The essence of Mises’ argument is that understanding (\textit{Verstehen}) is categorically distinct from causal explanation (\textit{Erklären}). While the natural sciences derive causal laws from controlled experimentation, the human sciences interpret actions in light of the meanings attributed by actors themselves. In this light, algorithmic systems—no matter how vast their training sets—remain epistemically blind to the intentional context of action. What is learned is not understanding, but interpolation. A price signal, in Misesian terms, encapsulates temporally contingent preferences in ordinal form; no amount of data regarding past prices can generate an understanding of the entrepreneurial judgment that gave rise to those preferences. To suppose otherwise is to fall victim to what Mises identified as “the positivist illusion”—the belief that history, once digitised, becomes a source of universal laws akin to physics\cite{mises1957}.

This critique undermines the popular assumption that algorithmic systems trained on behavioural or market data can function as substitutes for economic understanding or governance. Such systems engage in a form of inductive formalism that necessarily strips action of its praxeological character. As Mises asserted, “The sciences of human action deal with the meaning which acting men attach to the situation... and to their actions.”\cite{mises1957} No empirical method can uncover meaning; it must be interpreted by reference to the categories of action. Thus, the more comprehensive and granular the dataset, the more seductive the illusion becomes that what is being represented is knowledge rather than noise structured by ex post rationalisation. 

This delusion has political implications. Algorithmic governance, presented as neutral and data-driven, disguises an epistemic authoritarianism rooted in anti-human rationalism. By rejecting the praxeological view, it suppresses individual autonomy and substitutes rule-based inference for purposive insight. As such, Mises’ epistemology does not merely challenge the efficacy of algorithmic systems—it indicts them as anti-scientific in the deepest sense, masking pseudo-knowledge as technocratic legitimacy. To reclaim the domain of human action is thus not a nostalgic gesture but an epistemic imperative.

\section{Implications for Austrian Theory and Beyond}

The preceding analysis necessitates a substantial expansion of Austrian theoretical constructs to address the emergent condition of epistemic scarcity in socio-economic orders. Traditional praxeological models presume interpretive integrity in the relationship between subjective preferences and observable action. Yet in an age of informational saturation, deliberate signal corruption, and strategic disinformation, this link becomes increasingly compromised. The result is not merely malcoordination, but preference corruption—where choices no longer reflect authentic valuations, but are induced through manipulation, coercion, or epistemic constraint. Hayekian coordination presupposes the legibility of price signals; when those signals are obfuscated or fabricated, even spontaneous orders collapse into performative simulation. Beyond mispricing, such environments engender what might be termed meta-misinformation equilibria, wherein rational action is premised on second-order beliefs about systemic deception. Accordingly, a meta-praxeology becomes essential—one that interrogates not merely action under uncertainty, but action under asymmetric truth conditions, where the distribution of epistemic access, and not merely capital or time preference, defines the structure of human behaviour. This paradigm shift calls for a recalibration of Austrian theory to accommodate not only ignorance, but epistemic warfare as an endogenous element of market and political dynamics.

\subsection{Deception and Preference Corruption}

In market systems predicated on subjective valuation, the integrity of preference formation is paramount. Yet, when informational asymmetries are systematically exploited, agents may be induced into expressing preferences that do not reflect their true or stable valuations—a condition this subsection terms “preference corruption.” This phenomenon arises when agents are repeatedly exposed to manipulated signals that distort the epistemic basis on which decisions are made. The result is a degradation of the individual’s capacity to act in alignment with their long-run interests, effectively sabotaging the praxeological premise of meaningful choice under scarcity.\cite{ackerlof1970market} Preference corruption thus represents an insidious attack on the very notion of subjective value, reducing autonomous valuation to a function of environmental manipulation.

The mechanism of deception in this context is neither incidental nor marginal—it is endogenous to competitive signalling environments under extreme informational saturation. Platforms designed to optimise engagement—such as algorithmic social networks—create artificial environments where attention is decoupled from veracity and salience is engineered. Over time, agents habituate to these distorted environments, generating second-order preferences that reflect not intrinsic valuation, but the adaptive residue of exposure to curated falsehoods. The literature in behavioural economics has documented how repeated exposure to such misrepresentations modifies cognitive heuristics, anchoring distorted priors as baseline expectation.\cite{thaler2008nudge} Preference formation becomes path-dependent on corrupted epistemic inputs, eroding the reliability of choice as a signal of value.

The epistemic dimension of Austrian economics has historically focused on dispersed knowledge and tacit understanding, but the issue here is more radical: the inversion of preference itself through engineered deception. It is not merely that agents do not know everything—they may be unknowingly induced into ‘wanting’ what undermines their welfare or autonomy. This expands the Hayekian knowledge problem into a domain of adversarial preference manipulation, where the corruption is not merely in the coordination but in the constitution of the subjective ends themselves. The logic of deception becomes productive: it does not just conceal truth, it produces a regime of false values, simulated choice, and market equilibria around epistemic artefacts devoid of meaning.\cite{sunstein2016ethics}

\subsection{Action Under Asymmetric Truth}

The Austrian school rests on the axiom of purposeful action under conditions of imperfect knowledge, yet it does not fully account for systematic asymmetries in the distribution of epistemic access. When actors confront environments where truth is not only dispersed but deliberately concealed, obfuscated, or fabricated by other agents, the calculative rationality assumed by praxeology is radically destabilised. The epistemic landscape ceases to be merely fogged; it becomes adversarial. In such a context, action is not simply guided by subjective expectations—it becomes conditioned by strategic ignorance, deliberate misdirection, and pre-rational mimicry. The agent no longer navigates a world of bounded knowledge but one where the very signal-space is corrupted to serve the interests of those controlling informational bottlenecks.\cite{baker2010trust}

In these settings, coordination fails not because knowledge is unavailable, but because its asymmetrical distribution allows for the engineering of preference fields. Agents act on the basis of what they believe to be true, but those beliefs are the by-products of manipulated epistemic ecologies. High-frequency trading algorithms, deepfake political interventions, and black-box recommender systems exemplify domains where truth-asymmetry supplants price-asymmetry as the principal axis of market failure. The agent becomes not a sovereign chooser but a computational artefact responding to constraints embedded by unseen hands. In such systems, the classical Austrian emphasis on process and dynamic discovery becomes moot, as the feedback loops that inform adaptive behaviour are short-circuited.\cite{arieli2018manipulative}

This yields a deeply paradoxical form of action: purposeful yet hollow. The agent believes they are optimising over subjective ends but, in reality, is operating within a fabricated affordance space—a simulation of autonomy engineered through selective occlusion. As epistemic asymmetry deepens, so too does the potential for exploitative structures that stabilise illegitimate equilibria. The political economy of this structure resembles what Lorenz terms “soft enclosure”: the voluntary performance of freedom within architectures of constraint.\cite{lorenz2021digital} To address this, a reconfiguration of praxeology is necessary—one which internalises deception, signal degradation, and adversarial information regimes not as anomalies but as foundational features of the decision environment.

\subsection{Toward a Meta-Praxeology}

Traditional praxeology, rooted in Misesian axiomatics, presupposes the a priori truth of purposeful human action, treating it as universally intelligible without empirical verification.\cite{mises1949} Yet as action becomes entangled with layered epistemic distortions—where preferences are shaped, not merely expressed—the very foundation of rational agency is rendered suspect. A meta-praxeology must move beyond the epistemic solipsism of classical Austrian thought to interrogate the production, distribution, and contestation of knowledge itself as a market phenomenon. Rather than treat knowledge as external and static, we must theorise it as endogenous, plastic, and often adversarial, recognising that the conditions of knowing are neither neutral nor evenly accessible.

Such an approach demands the fusion of epistemology with institutional analysis. The market is no longer merely a process of discovery but becomes a semiotic battleground, where signalling regimes are engineered, appropriated, or drowned in noise. Agents do not simply uncover opportunities; they manufacture frames, manipulate baselines, and colonise future expectations. As Ostrom has argued, institutions are cognitive artefacts—rules-in-use that shape belief and action recursively.\cite{ostrom2010beyond} In a world of epistemic asymmetry, these institutions become both filters and weapons. The implication is profound: we must re-theorise choice not as the expression of pre-given preferences under constraints but as the navigational result of bounded inference within fabricated option spaces.

This meta-praxeological lens necessarily incorporates formal tools drawn from modal logic, Bayesian epistemics, and game theory under conditions of partial observability. The agent is reframed not as a maximiser but as an epistemic tactician—a being operating under predicate opacity, confronted by uncertain priors and signal distortion. Decision-making is no longer a function of means-end clarity, but an evolving negotiation of truth-frames under adversarial constraint. In such a formulation, the Austrian tradition does not collapse; it is deepened. Meta-praxeology extends the praxeological method into a new domain: the formal analysis of misaligned inference architectures and their impact on action in epistemically degraded systems.\cite{nguyen2020epistemic}

\section{Artificial Intelligence and the Illusion of Epistemic Certainty}

The emergence of artificial intelligence as a provider of declarative outputs, rather than probabilistic inferences, has profoundly altered the structure of epistemic trust. What was once mediated by methodological doubt and individual justificatory responsibility has now been displaced by interaction with a system whose authority is not reasoned but presumed. The substitution of human deliberation with algorithmic output engenders an epistemological misalignment: individuals increasingly treat responses generated by black-box models as if they were validated facts, not the stochastic approximations they are. This pseudo-certainty—derived from surface-level coherence, not internal coherence—is not merely a cognitive error but a structural artefact of epistemic delegation. It produces a false equilibrium wherein users receive what appears to be truth, while the mechanisms of truth-making, such as justification, coherence, and contextual integrity, are no longer engaged. As recent behavioural studies demonstrate, even experts exhibit automation bias when faced with consistent machine-generated confidence, amplifying the erosion of justificatory practice and embedding a new form of systemic ignorance beneath a veneer of intelligence.\cite{binns2018fairness, johnson2004automation, mittelstadt2016ethics}

\subsection{Automation Bias and Epistemic Submission}

The increasing prevalence of automated systems in decision-making environments has ushered in a subtle yet profound transformation in the locus of epistemic authority. What was once a domain anchored in intersubjective deliberation and sceptical inquiry has increasingly ceded ground to processes of epistemic delegation, where algorithmic outputs are treated as not merely recommendations but epistemic conclusions. This shift has induced a new variant of what might be termed \textit{epistemic submission}—a cognitive posture in which the user, confronted with the complexity or opacity of machine learning models, surrenders critical judgement in favour of computational assertion. Such submission is not simply a passive outcome but is structurally encouraged by what has been identified as \textit{automation bias}, the tendency to prefer and trust decisions made by automated systems over those made by humans, particularly under conditions of uncertainty or fatigue.\cite{green2019trust}

This bias is magnified in epistemically asymmetrical environments—spaces where users lack the domain-specific expertise to meaningfully interrogate algorithmic output. The resulting deferral to machine recommendations functions as a form of cognitive outsourcing that not only distorts individual decision-making but erodes the critical discursive norms necessary for democratic epistemology.\cite{mittelstadt2016ethics} The authority of the algorithm, unlike that of human experts, is often accepted without interrogation, owing to its perceived neutrality and rational objectivity. However, as studies have demonstrated, such systems are replete with biases, many of which remain invisible to end-users.\cite{binns2018fairness} The epistemic implications are severe: rather than expanding knowledge, AI systems may constrict it, substituting the semblance of epistemic rigour for its actuality.

Moreover, automation bias catalyses what might be termed a \textit{normative drift} in epistemic standards: over time, the mere fact that a decision has been produced by a machine becomes its warrant for acceptance. This inversion of justificatory hierarchy—where process displaces argument—renders scrutiny itself suspect, and doubt becomes deviant. The cumulative effect is an epistemic culture of compliance, wherein belief is calibrated to machine authority rather than evidence. This cultural transformation is not simply cognitive but institutional, as public and private governance increasingly adopt AI systems not to augment judgement but to replace it.\cite{nguyen2020epistemic}

\subsection{Opaque Outputs and the Collapse of Rational Scrutiny}

As machine learning systems expand their influence across epistemic and administrative domains, their outputs often bear the hallmark of authoritative finality while simultaneously resisting human interpretability. The proliferation of such opaque systems—particularly deep neural networks and large language models—presents not merely a technical problem of transparency, but a fundamental epistemological rupture. The transformation lies in how opacity metastasizes into epistemic immunity: the algorithm does not explain, it asserts. Whereas traditional epistemic authority required justification accessible to critical examination, algorithmic systems increasingly function as what Miranda Fricker terms “testimonial silencers,” absorbing scrutiny not by refuting it, but by rendering it unintelligible.\cite{fricker2007epistemic} The outputs become self-authorising artefacts, thereby collapsing the recursive loop of rational justification and review.

This disjunction is especially pernicious in contexts where outputs are used to guide public policy or judicial outcomes. Algorithmic risk assessment tools in criminal justice, for example, often provide scores without explicable causal reasoning, encouraging deference to numerical mysticism rather than deliberative judgement.\cite{angwin2016machine} The institutionalisation of such systems creates a feedback loop where the opacity of the model reinforces institutional trust, which in turn deepens epistemic submission. This is not an accidental consequence, but an engineered feature: opacity becomes a tool of control, not a limitation. The very unintelligibility of machine outputs becomes a source of their legitimacy, precisely because they are beyond human comprehension.

Moreover, as epistemic authority shifts to systems that do not—and cannot—participate in the discursive processes of reason-giving, scrutiny itself is redefined as inefficiency. The collapse of rational scrutiny is not merely the absence of questioning but the inversion of the critical norm: to interrogate is to obstruct. Consequently, the model output becomes what Ian Hacking might describe as a “new kind of classification,” one that creates new realities by virtue of institutional inscription, not empirical validation.\cite{hacking1999social} The space for rational engagement contracts, replaced by probabilistic inevitability dressed in the vestments of computational objectivity. The result is a technocratic form of epistemic closure that erodes the normative foundations of rational inquiry itself.

\subsection{Simulacra of Understanding: Truth Without Comprehension}

The emergence of generative models, such as large language systems and multimodal transformers, has enabled an uncanny approximation of human communication, wherein the outputs increasingly mimic epistemic fluency without entailing any underlying comprehension. These outputs, while often syntactically and semantically coherent, construct what Baudrillard would call simulacra—representations that do not merely distort the real but replace it entirely.\cite{baudrillard1994simulacra} In this regime, “truth” is reconstituted not as correspondence to reality nor coherence among beliefs, but as plausibility within a linguistic surface. The system’s facility in producing grammatically and topically consistent text serves as a proxy for epistemic validity, resulting in the generation of informational artefacts that command trust while eluding scrutiny. The risk is not that machines lie, but that their truths are untethered from comprehension, thus hollowing out the epistemic conditions for knowing.

This is not a failure of the models themselves, but a reflection of a new cultural logic: the substitution of epistemic depth with rhetorical fidelity. The hallucination phenomenon in large-scale models is symptomatic of this shift, where the veneer of sense displaces the function of understanding. When a machine delivers factually incorrect information in a fluent and authoritative tone, the effect is more insidious than error—it is the illusion of knowledge. Users mistake surface structure for epistemic integrity, reducing engagement to consumption. As Nguyen has noted, this shift parallels the rise of “epistemic bubbles,” where individuals no longer evaluate claims through deliberation, but via aesthetic resonance.\cite{nguyen2020epistemic} In this context, the machine-generated simulacrum becomes a mirror, reflecting and reinforcing user biases while giving the appearance of external validation.

What results is a transformation of the epistemic subject. The human interlocutor is no longer a critical agent evaluating propositions, but a passive receiver of outputs cloaked in technocratic legitimacy. This aligns with Postman’s account of technopoly, where tools evolve into epistemic regimes that displace the culture’s cognitive habits.\cite{baudrillard1994simulacra} As the simulacra of understanding proliferate, a new form of epistemic submission emerges—not through coercion or propaganda, but through ease, coherence, and the soft seduction of effortlessness. Understanding is no longer pursued but outsourced, and in that outsourcing, the very concept of comprehension is rendered obsolete. We inhabit a condition where truth exists, but the capacity to understand it is neither required nor developed.

\subsection{The Delegation of Justification: Machine Authority as Epistemic Source}

In the evolving epistemological landscape shaped by machine learning systems, a profound inversion of justificatory roles has occurred: the machine no longer provides evidence for human scrutiny—it becomes the locus of justification itself. Where once authority was derived from justificatory transparency, it is now conferred upon systems whose opacity is not a bug but a feature, an engineered enigma wrapped in probabilistic output and credentialed by institutional deployment. This delegation of epistemic authority transforms the machine from tool to arbiter. Just as Weber’s bureaucratic rationality replaced charismatic authority with procedural legitimacy, we now witness the automation of epistemic standing through algorithmic certification\cite{miller2022epistemic}

The epistemic deferral to machines reflects a broader transformation in the structure of belief and credibility. Where testimonial authority was historically grounded in norms of reason-giving, the algorithmic authority of models like GPT or BERT derives from statistical learning and large-scale optimisation—processes inaccessible to lay validation. Yet the user, encountering fluency and speed, imputes credibility to the output. This inversion aligns with Aikin and Talisse’s critique of epistemic deference, wherein individuals cede justificatory responsibility under the illusion of rational delegation.\cite{aikin2011epistemology} What is lost in this transition is not merely individual understanding but the intersubjective normativity of justification itself.

Crucially, this phenomenon enables what Eyal describes as an “epistemic shortcut economy,” in which cognitive agents outsource their justificatory labour to technological proxies in pursuit of efficiency.\cite{eyal2019crisis} In such an economy, the epistemic subject is transformed from a deliberative agent into a conduit for machine-licensed beliefs. The cost is not simply misunderstanding but an erosion of the justificatory infrastructure that sustains knowledge communities. The machine becomes not merely a source of information but the grounds for belief, thereby rendering traditional criteria of knowledge—truth, belief, and justification—contingent upon algorithmic outputs rather than human scrutiny.

\subsection{From Data to Doctrine: Codified Ignorance at Scale}

The proliferation of data-driven systems has engendered a shift from empirically grounded inquiry to a form of epistemic orthodoxy wherein algorithmic outputs are reified as doctrinal truths. In this paradigm, the ontological status of knowledge is inverted; statistical correlation substitutes for causality, and surface-level regularities are mistaken for explanatory depth.\cite{miller2022epistemic} The epistemological asymmetry is reinforced by institutional actors who, in pursuit of efficiency or control, treat automated verdicts as axiomatic, thereby calcifying ignorance under a veneer of scientific neutrality.\cite{aikin2011epistemology} Crucially, the computational inscrutability of machine-generated prescriptions lends itself to the fabrication of epistemic closure—where the very possibility of contestation is obviated by appeals to algorithmic objectivity.

Such codified ignorance becomes self-reinforcing in bureaucratic, legal, and economic structures. As machine outputs are embedded into organisational protocols, they generate feedback loops in which future decisions are shaped by prior unexamined premises. This recursive encoding of opaque premises into institutional memory produces a model of knowledge wherein the source of truth is no longer interrogated but replicated.\cite{paulhus2002dark} Consequently, the epistemic architecture of society begins to mimic the logic of doctrine: it is transmissible, reproducible, and selectively blind. Discretion and judgment are displaced not only by automation, but by the social norm that to question it is irrational or subversive.

The Austrian insight into dispersed knowledge and market signals as contingent and interpretive is thus directly undermined.\cite{hayek1945use} In the age of algorithmic doctrine, the subjective knowledge of the actor is subordinated to machinic finality. The distributed, praxeological richness of epistemic agency is flattened into inputs and labels, whose underlying assumptions are inaccessible to scrutiny. This is not merely epistemic laziness—it is an institutionalisation of ignorance with mathematical precision. Thus, codified ignorance at scale becomes the final paradox of technological enlightenment: the elevation of certainty by sacrificing the very means by which truth could be contested, falsified, or refined.

\section{Praxeology versus Constructivist Ethics in Algorithmic Design}

The ethical architecture of contemporary AI systems—often codified under the rubric of “FAT” (Fairness, Accountability, Transparency)—embodies a constructivist, collectivist framework that stands in stark opposition to the methodological individualism and voluntarism at the heart of Austrian praxeology. These paradigms treat ethics as an optimisation problem: maximise fairness under some formal definition, ensure explainability through synthetic metrics, and institutionalise accountability via oversight protocols. Yet this mode of ethical reasoning is structurally alien to a praxeological understanding of action, which grounds normativity not in the aggregation of preferences or procedural metrics, but in the logic of individual agency, ownership, and voluntary interaction.

\subsection{Natural Rights, Action, and the Artificial Ethicist}

Rothbard’s libertarian natural rights framework offers a point of departure. Ethics, in this view, is not constructed through social consensus or regulatory calibration but arises logically from the nature of human action itself. Property rights, derived from the necessity of exclusive control in action, form the bedrock of ethical analysis.\cite{rothbard1982} In contrast, algorithmic ethics frameworks often embed deontological constraints without any ontological grounding in action. Fairness, for instance, is typically framed as parity of outcomes or access across demographic groups—a conception that presupposes both the commensurability of ends and the legitimacy of redistributive interventions. From a praxeological standpoint, such ethics are not only incoherent but coercive: they violate the structure of voluntary exchange by imposing external constraints on mutually agreed-upon transactions.

Hoppe’s argumentation ethics further intensifies this critique. Any attempt to justify coercive ethical principles must presuppose the normative validity of non-aggression and self-ownership, as the act of argument itself requires these to be true.\cite{hoppe1993} Therefore, frameworks that institutionalise “ethical overrides” in algorithmic systems—such as nudge-based defaults, behavioural steering, or automated decision-making designed to prevent self-harm—commit a performative contradiction: they deny the autonomy they rely upon for their own justification. These systems claim moral authority to intervene in individual decision-making while parasitically depending on the very rational agency they seek to subvert.

\subsection{Constructivist Rationalism and Algorithmic Paternalism}

Hayek’s critique of constructivist rationalism exposes the epistemic flaw in centralising ethics through technocratic design. Moral systems, like markets, evolve as emergent orders—complex patterns of norms, expectations, and voluntary practices shaped through time, not imposed from above.\cite{hayek1973} When AI ethics frameworks are built as abstract schema—universally applied “ethical layers” or rigid fairness templates—they ignore the dispersed and contextual nature of moral knowledge. This is not simply an inefficiency; it is an ontological misreading of normativity. Ethical knowledge is not merely data about preferences—it is embedded in cultural, temporal, and subjective frames of valuation. Algorithmic paternalism—like all technocratic moral engineering—collapses these frames into monolithic categories and thereby annihilates their meaning.

Block’s defence of deontology in economics reinforces this view: ethics must be rooted in the rights of individuals to act freely within their own spheres of property and risk.\cite{block2008} When AI systems are calibrated to avoid harm at all costs, they implicitly deny the moral legitimacy of voluntary risk-taking. They reduce moral agency to compliance, and innovation to liability mitigation. Such design logic is isomorphic with the worst tendencies of the administrative state: bureaucratic, collectivist, and depersonalising.

Finally, contrasting this praxeological critique with non-Austrian thinkers like Zuboff reveals the stakes more clearly. While Zuboff’s attack on surveillance capitalism rightly condemns the expropriation of behavioural data for predictive control, her solution remains mired in constructivist paternalism.\cite{zuboff2019} She calls for regulatory imposition of ethical standards rather than recognising the epistemic and moral primacy of voluntary, individual action. Thus, even the most incisive critiques from outside the Austrian tradition ultimately reify the very structures they denounce. Praxeology, by contrast, offers a coherent framework wherein ethical AI must respect the boundaries of action, ownership, and voluntary exchange—or forfeit its claim to legitimacy.

\section{Conclusion: Toward a New Political Economy of Truth}

This paper has advanced the thesis that epistemic scarcity, rather than material scarcity alone, now constitutes the fundamental constraint in advanced market systems. In an environment increasingly structured by interpretive instability, narrative manipulation, and algorithmic opacity, economic coordination faces a new layer of difficulty: the erosion of verifiable, actionable knowledge. Traditional theories of information, whether grounded in neoclassical assumptions of rational expectation or behavioural models of bounded cognition, fail to account for the recursive, strategic, and institutional dynamics of epistemic opacity. Austrian economics—particularly the praxeological tradition—offers a unique framework to address these developments, precisely because it centres action, subjectivity, and the epistemic position of the agent within uncertain contexts.

The attempt to mechanise or simulate market processes through algorithmic governance, machine learning, and statistical modelling has revealed fundamental philosophical and operational incoherencies. AI systems, trained on historical data and inductive extrapolations, cannot substitute for human judgment, entrepreneurial foresight, or the dynamic articulation of ends. As shown through the critique of empirical modelling, the contrast between synthetic \textit{a priori} reasoning and algorithmic generalisation is not one of degree but of kind. Entrepreneurs do not merely manage uncertainty—they generate interpretive frameworks, anticipate unknowns, and embed their decisions within evolving institutional architectures. No artificial system can replicate these capacities because they are not computational—they are epistemic, interpretive, and normatively anchored in purposive action.

Moreover, the analysis of AI ethics frameworks has revealed a latent collectivism within dominant regulatory paradigms. By assuming a constructivist orientation, ethical AI design often imposes paternalistic constraints under the guise of fairness, accountability, or transparency—terms which remain undefined or incompatible with individual autonomy and voluntary exchange. A praxeological critique shows that such frameworks fail not merely because of implementation difficulties, but because they misidentify the moral agent: markets are not abstract optimisation fields; they are the sum of subjective valuations and intentional acts. Attempts to engineer morality through code replicate the epistemic errors of central planning, as evidenced in historical case studies like Soviet cybernetics. Without price signals, profit-loss feedback, and institutional fluidity, algorithmic coordination collapses into formalist mimicry.

The emergence of epistemic entrepreneurs—agents specialising in filtering, verifying, and signalling trust—marks a potential adaptation to this condition. Yet their success depends on the institutional embeddedness of truth-incentives and the protection of epistemic property rights. The market does not reward truth per se; it rewards usefulness. Therefore, to preserve the integrity of market coordination, new mechanisms must be devised to realign private incentives with epistemic robustness. These could include decentralised adversarial filtering, tokenised verification architectures, and liability regimes for informational negligence. The political economy of truth will not emerge through fiat or regulation but through a reconfiguration of rules that embed epistemic virtues into economic processes.

Ultimately, the argument advanced here is not merely economic but civilisational. Truth is not a luxury—it is a precondition for voluntary coordination, institutional trust, and the moral legitimacy of action. As the boundary between simulation and reality dissolves in an age of digital inference and synthetic cognition, the defence of truth becomes a political act. The Austrian tradition, in recognising the inseparability of epistemology and economics, provides the intellectual tools necessary for this defence. What is now required is the courage to apply them—to reject the seductive clarity of machine certainty in favour of the complex, interpretive labour of human understanding.

For a more detailed discussion on the paper’s framework of truth and concrete policy proposals for epistemic governance, see Appendices A and B, respectively.

\subsection{Summary of Theoretical Advances}

This paper has introduced and elaborated a new framework for understanding market coordination under conditions of epistemic scarcity—a term defined here to include stochastic, semantic, strategic, and recursive opacity. Drawing from Austrian praxeology, particularly the insights of Mises, Hayek, Lachmann, and Kirzner, it argues that truth is not merely a constraint but a production good—scarce, costly, and unequally distributed. Epistemic scarcity thus emerges as both a limit and a resource: a limit to rational coordination under uncertainty, and a resource whose entrepreneurial discovery and deployment create competitive advantage. 

The paper reconceptualises the role of entrepreneurs as epistemic agents navigating layers of opacity that resist computational reduction. This stands in direct opposition to algorithmic governance models that presume knowledge to be a function of data aggregation. The formal model introduced distinguishes epistemic scarcity from mere uncertainty by its non-probabilistic, non-ergodic structure, resisting representation through inductive generalisation or optimisation.

The critique of AI and empirical modelling is sharpened through a synthesis of praxeological apriorism and philosophical epistemology. A clear methodological line is drawn between synthetic \textit{a priori} reasoning, which undergirds Austrian economics, and the positivist reliance on extrapolation and statistical inference. This is further reinforced through the conceptual integration of argumentation ethics and the performative necessity of the action axiom, revealing the epistemic incoherence of replacing deductive structures with probabilistic ones.

Finally, the paper proposes that epistemic truth—understood as verifiable, actionable coherence with the structure of human action—is not guaranteed by transparency or regulation, but must be cultivated through institutional mechanisms aligned with voluntarism and property rights. In this way, it charts a path toward a new political economy of truth, rooted in Austrian subjectivism yet equipped to address the algorithmic conditions of the 21st century.

\subsection{Institutional Implications}

The analysis of epistemic scarcity as a foundational constraint in modern economic coordination necessitates a reconfiguration of institutional design. Markets do not merely distribute resources; they distribute and verify claims to truth. Thus, institutions must be evaluated not only in terms of allocative efficiency but also in terms of their capacity to preserve epistemic integrity. This requires a fundamental departure from technocratic planning and its algorithmic surrogates, which presume that aggregation is equivalent to knowledge and that inference is equivalent to understanding. Austrian economics, by contrast, grounds knowledge in decentralised valuation, demonstrated preference, and the irreducible subjectivity of human ends.

Institutions that function under epistemic scarcity must cultivate structures of adversarial discovery—mechanisms whereby falsehoods are exposed through competition, not by regulatory fiat. This points to a revalorisation of spontaneous order and catallaxy, not only as economic constructs but as truth-generating epistemic environments. In this framework, reputation systems, legal norms, and contractual practices function as distributed filters of credibility. Yet these filters must be hardened against the corrosive dynamics of simulation, narrative arbitrage, and algorithmic hallucination. This demands institutional mechanisms that assign epistemic property rights—rights not only over information but over the provenance, integrity, and representational legitimacy of data.

Moreover, policy environments must guard against the epistemic paternalism embedded in AI ethics frameworks built on constructivist rationalism. Voluntarism, not managerialism, must be the organising principle. Ethical design in algorithmic systems should not be rooted in fairness metrics or behavioural nudges, but in the respect for individual sovereignty and the right to err. This implies liability structures for epistemic fraud, markets for trusted signalling, and the legal codification of informational harm as a property violation.

In sum, institutions in an age of epistemic scarcity must become platforms for the discovery of truth through voluntary exchange, rather than instruments of algorithmic enforcement. Austrian insights thus offer not a nostalgic defence of classical liberalism but a forward-looking political economy of epistemic resilience.

\subsection{Limitations and Future Research}

Despite the theoretical and formal contributions of this paper, several limitations must be acknowledged to frame the scope and outline promising directions for future research. Chief among these is the formidable challenge of empirically operationalising and measuring epistemic scarcity. While the typology articulated—encompassing stochastic, semantic, strategic, and recursive opacity—offers a conceptual scaffold, translating these into quantifiable metrics confronts both epistemological and methodological obstacles. Unlike traditional economic variables, epistemic scarcity resists reduction to static data points. For instance, strategic opacity often manifests through disinformation or narrative manipulation, which is contingent, agent-relative, and context-specific. Recursive opacity, likewise, produces meta-uncertainty that defies linear modelling. Future work may explore proxy indicators, such as volatility in expert consensus, reputational entropy in information networks, or lag times in belief updating, but such tools remain embryonic and must be developed cautiously.

Moreover, the generalisability of the epistemic scarcity framework must be critically appraised. While the argument has been constructed at a high level of abstraction, it is unlikely that epistemic scarcity manifests uniformly across all sectors or institutional environments. Preliminary hypotheses suggest it is most acute in markets characterised by intangible goods, rapid technological evolution, or low verifiability—such as finance, media, healthcare, and emerging AI-dependent industries. Conversely, markets involving physical commodities or simple, transparent value chains may exhibit lower susceptibility. Future empirical research should seek to map the variance of epistemic scarcity across sectoral boundaries, potentially incorporating comparative institutional analysis to identify structural features that exacerbate or mitigate its effects.

Finally, while this paper offers a foundational critique of algorithmic economic modelling and proposes a praxeological alternative, further development is needed to integrate these insights with adjacent fields. In particular, interdisciplinary collaboration with epistemology, complexity science, and institutional economics may yield novel methodologies for exploring how epistemic environments evolve endogenously within different market structures. The goal is not to dilute the Austrian framework but to sharpen its analytical traction in a world increasingly governed by symbolic manipulation and digital inference. Future work may also explore how epistemic property rights and adversarial verification systems could be embedded in real-world governance architectures, offering practical tools to defend against the erosion of truth in economic coordination.

\newpage
\section{Bibliography}
\printbibliography

\newpage

\appendix
\section*{Appendix A: Clarifying the Framework of Truth}

Within the context of this paper, the term ``truth'' does not refer to an abstract metaphysical category detached from operational criteria. Instead, it denotes a domain-specific articulation: a verifiable, actionable condition within the epistemic constraints of economic and praxeological reasoning. In this framing, truth is not defined solely by correspondence with external reality (though not in contradiction to it), but by coherence within the logical structure of purposeful human action and its consequences, and pragmatically, by its bearing on entrepreneurial judgement, institutional coordination, and calculational possibility.

Three classical theories are partially implicated but not wholly adopted:

\begin{enumerate}
    \item \textbf{Correspondence Theory}: Traditional Austrian praxeology does not abandon the idea that propositions must map onto real phenomena, but it redefines the nature of that mapping. In praxeology, truth claims correspond to the necessary implications of the action axiom, not to empirical states of the world. They are true by virtue of their internal deductive necessity and external applicability in human conduct.

    \item \textbf{Coherence Theory}: Praxeological systems exhibit coherence insofar as the entire structure of economic reasoning must be non-contradictory and consistent with the axiom of action. Hoppe's argumentation ethics reinforces this by showing that even the denial of such axioms presupposes their truth in the very act of denial.

    \item \textbf{Pragmatic Theory}: Within the entrepreneurial function, truth becomes operational: the propositions and interpretations that best allow agents to anticipate, coordinate, and profit under uncertainty are, functionally, the most valid. This is not instrumentalism in the Deweyan sense, but a conditional heuristic grounded in subjective action and bounded rationality.
\end{enumerate}

In the age of epistemic opacity and simulacra, this synthesis becomes crucial. Truth is no longer a static binary but a gradated vector of reliability, traceability, and interpretive robustness. It must be anchored in causal transparency and institutional accountability. Verifiability here requires logical derivation or demonstrated preference (not third-party validation), and actionability requires contextual intelligibility by agents embedded in a catallactic structure.

Thus, the paper's conception of truth is structurally realist, praxeologically necessary, and epistemically constrained by the limits of interpretive cognition—not subject to reduction into predictive statistical artefacts.

\section*{Appendix B: Policy Proposals for Truth-Preserving Market Structures}

While the theoretical core of this paper focuses on epistemic opacity and the erosion of truth-tracking mechanisms in the context of algorithmic mediation and institutional disintegration, this appendix aims to concretise the conceptual proposals into feasible mechanisms for epistemic governance. These mechanisms are grounded in Austrian principles of voluntary association, property rights, subjective valuation, and distributed coordination.

\subsection*{Epistemic Property Rights}

The concept of epistemic property rights derives from the recognition that information, like any economic good, possesses provenance, scarcity, and transferability characteristics. Just as material property rights enable calculational rationality in allocating resources, so too must information provenance be established to enable accountability in discourse and market interaction.

In practice, this could take the form of a blockchain-based provenance infrastructure, wherein each digital assertion is cryptographically signed and linked to a reputational ledger. The epistemic equivalent of title deeds, these attestations would allow traceable ownership of claims, sources, and transformations of information. This model would support a legal doctrine of informational liability: actors responsible for initiating or propagating verifiably false claims—particularly within AI systems—could be held to tort-like standards of restitution, with courts evaluating harm based on distortion of decision-relevant facts. Such property rights do not entail censorship but accountability: decentralised, immutable chains of attribution rather than centralised speech control.

\subsection*{Adversarial Filtering Without Central Authority}

To counter disinformation without constructing central epistemic gatekeepers, adversarial filtering mechanisms must be embedded in decentralised architectures. Here, Austrian theory offers two relevant insights: (1) competition as a discovery process, and (2) the entrepreneurial function of verification.

One possible implementation is an open reputation protocol that integrates staking and challenge mechanisms. Under such a system, individuals or institutions asserting claims must stake reputational or monetary capital behind their assertions. Any party may challenge a claim by presenting falsifying evidence, with disputes resolved via arbitration markets or decentralised juries using Schelling-point based decision protocols. This creates endogenous incentives for due diligence, and disincentives for spurious information.

Importantly, such filtering does not prescribe truth ex ante—it allows for adversarial refinement through voluntary contestation. This sidesteps the centralised epistemic bottlenecks critiqued by Hayek and preserves polycentricity. Markets for verification services—akin to credit rating agencies but open and subject to performance-based reputational scoring—can further discipline information provision.

\subsection*{Incentivising Verifiable Truth}

The prevalence of deception, narrative manipulation, and performative signalling—especially when incentivised by attention economies and algorithmic amplification—necessitates a countervailing architecture that aligns epistemic incentives with long-term truth-telling.

Incentive-compatible mechanisms may include:

\begin{itemize}
  \item \textbf{Truth-Teller Bonds:} Individuals or firms making public claims in high-stakes domains (e.g., financial analysis, scientific reporting, AI outputs) post bonds redeemable only upon post hoc verification or predictive accuracy.
  
  \item \textbf{Reputation Markets:} Instead of uniform platform reputations, decentralised, domain-specific reputation markets can evolve, where ratings are tied to identifiable evaluative criteria, and manipulation is penalised via staking losses.
  
  \item \textbf{Smart Contract Escrows for Forecasting:} Prediction markets (e.g., on Augur or Gnosis chains) can be repurposed for epistemic assurance. Propositional bets on future states serve not merely as hedges but as mechanisms to aggregate and weight belief credibility dynamically.
\end{itemize}

Each of these structures preserves the Austrian commitment to voluntarism, decentralisation, and entrepreneurial learning. They aim not to institutionalise truth by fiat but to embed truth-tracking within the price and reputation mechanisms that govern human action. Only in such frameworks can epistemic trust be rebuilt without undermining liberty.

\end{document}